\documentclass[twocolumn]{aastex631}
\usepackage{outlines}
\usepackage{enumitem}
\usepackage{float}
\usepackage{amsmath}
\usepackage{siunitx}
\usepackage{mathtools}
\usepackage{hyperref}
\usepackage{graphicx}
\usepackage{color}
\usepackage{soul}
\let\bfseries\mdseries

\begin{document}

\newcommand{\vdag}{(v)^\dagger}

\newcommand{\figr}[1]{Figure~\ref{fig:#1}}
\newcommand{\secr}[1]{Section~\ref{sec:#1}}
\newcommand{\tabr}[1]{\mbox{Table~\ref{tab:#1}}}




\shorttitle{Effects of Faculae}
\shortauthors{Li \& Basri}




\title{Do Faculae Affect Autocorrelation Rotation Periods in \textbf{Sun-like} Stars?}


\author{Canis Li}
\affil{Valley Christian High School, San Jose, CA 95111}
\author{Gibor Basri}

\affil{Department of Astronomy, University of California, Berkeley, CA 94720}

\email{gbbasri@berkeley.edu} 


\date{January 21, 2024}



\begin{abstract}

\textbf{Rotational periods derived from autocorrelation \textbf{(ACF)} techniques on stars photometrically similar to the Sun in \textit{Kepler} data have proven difficult to reliably determine}. We investigate various instrumental and astrophysical factors affecting the accuracy of these measurements, including the effects of observational windows \textbf{and noise, stellar activity and inclination}, spectral passbands, and the separate normalization of contiguous segments. \textbf{We validate that the flux variations due to faculae are very periodic, but} starspots are the dominant source of \textbf{bolometric and visible} differential variability in \textbf{Sun-like stars on rotational timescales}. We quantify how much \textbf{stronger the relative} contribution of faculae would have to be to render \textbf{Sun-like} light curves periodic enough to reliably measure with autocorrelation methods. We also quantify how long starspot lifetimes need to be to render pure spot light curves periodic enough. \textbf{In general, longer observational windows yield more accurate ACF measurements, even when faculae are not present. Due to the enhancement of the relative contribution of faculae, observing stars with intermediate inclinations, during activity minima, and/or through bluer passbands has the effect of strengthening the periodicity of the light curve.} We search for other manifestations of faculae in broadband photometry of \textbf{Sun-like} stars and conclude that without absolute flux measurements or restriction to shorter wavelength passbands, differential light curves are uninformative about faculae. 

\end{abstract}


\section{Introduction}

Missions such as CoRoT (\textbf{Convection, Rotation and planetary Transits}; \citealt{Auv09}) and \textit{Kepler} \citep{Bor10} have provided a wealth of high-precision broadband photometric data on stars over long time periods; a boon for stellar researchers as well as studies of exoplanets. Current missions such as TESS (\textbf{Transiting Exoplanet Survey Satellite}; \citealt{TESS}) are continuing this stream of data, and \textbf{PLATO (PLAnetary Transits and Oscillations of stars;} \citealt{PLATO}) will soon provide a large new sample. Stellar brightness variability on timescales of days to months is dominated by the rotational modulation of starspots and faculae on the surface of stars. Both starspots and faculae are photospheric phenomena brought about by magnetic fields on the stellar surface. Starspots are darker than the quiet photosphere because strong magnetic fields inhibit energy transfer via convection. The formation of faculae in \textbf{Sun-like} stars is more subtle. They are essentially optical depth effects caused by the sweeping of magnetic fields to the edges of granules (convective cells), and the subsequent partial substitution of magnetic for gas pressure there. Because the gas opacity is reduced, one can see further into the hot cell interiors, making for a brighter optical surface \citep{dePont2006}. Faculae are better seen away from disk center because the lines of sight intersect the cell edges more fully near the limb. Active regions can also be made brighter by magnetic heating in stronger flux tubes. 

High-resolution intensity and magnetogram images of the Sun allow investigations of individual spots and faculae on the solar surface. But our ability to study starspots and stellar faculae is limited by the lack of spatial resolution when viewing distant stars. Absolute photometry (mostly ground-based) reveals that stars with similar activity levels as the Sun exhibit its tendency to be brighter due to faculae when also covered with more spots (during activity cycle maxima). Such stars are called ``faculae-dominated". On the other hand, more rapidly rotating stars have generally more magnetic activity and tend to be darker near activity maxima (as determined, for example, by CaII measurements of the chromosphere). This is presumably because their absolute photometry is dominated by starspots (\citealt{Lock2007, Hall2009}). These are called ``spot-dominated"; both designations refer to the behavior of absolute brightness relative to magnetic activity. \textbf{Although the Sun is faculae-dominated as defined above, its TSI variations on \textit{rotational} timescales are generally dominated by spots when viewed from the ecliptic \citep{Shapiro2016}. It turns out that whether faculae or spots dominate the rotational brightness modulations depends on both the viewing inclination and the wavelength passband; \citet{Shapiro2016} found that for moderate inclinations faculae tend to dominate the total solar irradiance (TSI) variations, while spots dominate at higher inclinations such as that of the Sun.}

\textbf{In the context of this paper ``Sun-like" is taken to mean that the stellar light curve shares certain characteristics with the observed solar light curves. These include a weak periodicity induced primarily by the dominance of spots which live typically not much more than one rotation (and often less). By ``weak" we mean difficult to detect using autocorrelation methods with observing timescales on the order of a year. This class of stars also show relatively low amplitudes of photometric variability compared with the variability seen in most of the stars whose periods were determined in the first analyses of \textit{Kepler} light curves. Finally, ``Sun-like" stars are presumed to have rotation periods roughly three weeks or more, including those longer than the Sun's. In practice, this paper is restricted to light curves that are actual or model versions of solar data, with the exception of models that test the effect of starspot lifetimes.}

Most of the variable signal from \textbf{Sun-like stars} on the timescale of days consists of dips in the differential light curve due to starspots, but we know from the Sun that there must also be a positive (brightening) signal from faculae. Contemporary models (e.g., \citealt{SATIRE, Johnson2021}) depict solar and stellar variability light curves as composed of a facular signal and a starspot signal superposed to form a net signal. The signature of starspot variability in the light curve arises from rotational disk passages of starspot groups, coupled with starspot evolution. The same is true for faculae, but their spatial and temporal distributions are different. \textbf{A fuller discussion of the effects of magnetic fields on photometry can be found in \citet{BasriBook}.}

In the case of slowly rotating (periods of about \textbf{three weeks} or more) Sun-like stars, spots usually live from days to weeks \citep{Solanki2003}. If \textbf{spot} evolution has changed the global spot distribution too much by the second disk passage, the spot signal is increasingly aperiodic. The solar facular signal is generally more periodic than the spot signal due to the tendency of faculae to be more long-lived and more widely spread across the solar surface \citep{Chapman1997}. \textbf{However, the spots generally induce larger brightness variations than faculae on rotational timescales \citep{Shapiro2016}, making Sun-like rotation periods more difficult to determine}. For stars more magnetically active than the Sun, such as young rapidly rotating stars, a strongly periodic rotation signal can be seen in their light curves because their starspots live for many rotational periods \citep{Basri2022}. Such spots can be very long-lived and often concentrated more poleward, and can begin to crowd out the faculae \citep{Johnson2021}. It is worth noting that faculae in stars with temperatures much different from the Sun may have different contrasts relative to the quiet photosphere, and even be darker than it in very cool stars \citep{Norris2023}.

Stellar rotation periods are desirable to know because of the insights they provide into gyrochronology as well as the dynamo mechanisms driving stellar magnetic activity. \textbf{Until being surpassed by Gaia\footnote{\textbf{The periods found by Gaia and TESS are also mostly restricted to a week or less because of the observing strategies of these missions.}}}, the most abundant source of stellar rotation periods has been the \textit{Kepler} space mission.  A plethora of timeseries analysis techniques have been applied to identify the rotation period of stars from their rotational light curve modulations. These include the Lomb-Scargle periodogram, autocorrelation functions, and wavelet analysis (e.g., \citet{Aigrain2015, Santos2021}). In a highly cited paper \citet{MMA2014} utilized autocorrelation functions (ACF) to derive rotational periods from \textit{Kepler} differential photometry, detecting the rotation periods of 34,030 out of the 133,030 main-sequence \textit{Kepler} targets. 

Recently, \citet{Reinhold2021} investigated the effect of inclination and metallicity on period detectability from differential light curves, concluding that \textbf{solar-like stars} are underrepresented in the period catalog of \citet{MMA2014} because of the low accuracy of \textbf{period measurements} from their light curves. \textbf{The effects of inclination and metallicity on Sun-like brightness variations were previously analyzed by \citet{Nemec2020} and \citet{Witzke2018}, respectively.} \citet{Basri2022} also found that solar-type stars are significantly underrepresented among main sequence stars with detected rotational periods, and that the ACF often finds incorrect (\textbf{sometimes half}) periods for these difficult cases, partly because of the \textit{Kepler} reduction pipeline. Recently a new Gradient of the Power Spectrum (GPS) method \citep{Shapiro2020} has been shown to work better for solar-type stars where the other methods struggle \citep{Reinhold2023} (with a small subset of doubled periods). This method apparently works better because it does not try to detect periodicity directly but depends more on the duration of features in the differential light curve (which is related to the rotation rate). Another possibly improved method for these cases is \textbf{based on Gaussian process regression} \citep{Angus2018}. Here we concentrate on the accuracy of \textbf{ACF-based methods} in finding periods from differential light curves, and our results may be applicable to any methods that rely directly on the periodicity of the signal.

Attempts have also been made with \textit{Kepler} data to characterize the absolute brightness variations of faculae on the timescale of activity cycles, although \textit{Kepler} collected rather little absolute photometry. \citet{Montet2017} utilized the monthly \textit{Kepler} full field images to measure the absolute brightness of stars and search for activity cycles. Their sample of facular-dominated stars has the right general behavior, but contains exceptions to the dependence on the rotation period expected by the transition from spot to facular domination as a star becomes older and rotates more slowly \citep{Shapiro2014}. In particular, there is a set of very rapid rotators that show a \textbf{facular dominance} (brightness increases with activity level) despite the fact they should be very spot-dominated \citep{Basri2018}. This result was confirmed in Gaia data by \citet{diStef2023}. 

\citet{Basri2018} also found that in the solar case, the increase in absolute brightness from faculae cannot be discerned in \textit{Kepler} differential light curves because the ``clean star" continuum level is undetermined. \citet{Johnson2021} show that the range in differential variability ($R_\mathrm{var}$) is not affected much by increasing the facular to spot area ratio, and also discuss changes in absolute brightness as this ratio varies. Very little is known empirically to date about how faculae manifest in the \textit{differential} (as opposed to absolute) brightness rotational modulations of stars.

This paper advances the analysis of the information content in stellar differential light curves by looking separately at the effects of faculae, spots, and other observational or astrophysical factors. It concentrates on the detectability of the stellar rotation period using ACF methods. We start with the effects of observational factors, then investigate the role of faculae in yielding a reliable ACF rotation period for \textbf{Sun-like} stars. For this we utilize collections of model light curves with known spot and facular components and known periods to assess the statistical accuracy of ACF period determinations. Finally, we search (unsuccessfully) for clearly identifiable features of faculae in differential light curves other than their tendency to make the net signal more periodic.

\section{Ingredients of the Analysis}

In this section we describe the various sorts of analysis we undertake and the light curves we conduct them on, while leaving the results of those analyses to the following section.

\subsection{Time \textbf{S}eries \textbf{A}nalyses} \label{sec:timeseriesanalysis}

\begin{figure*}[htb!]
\gridline{\fig{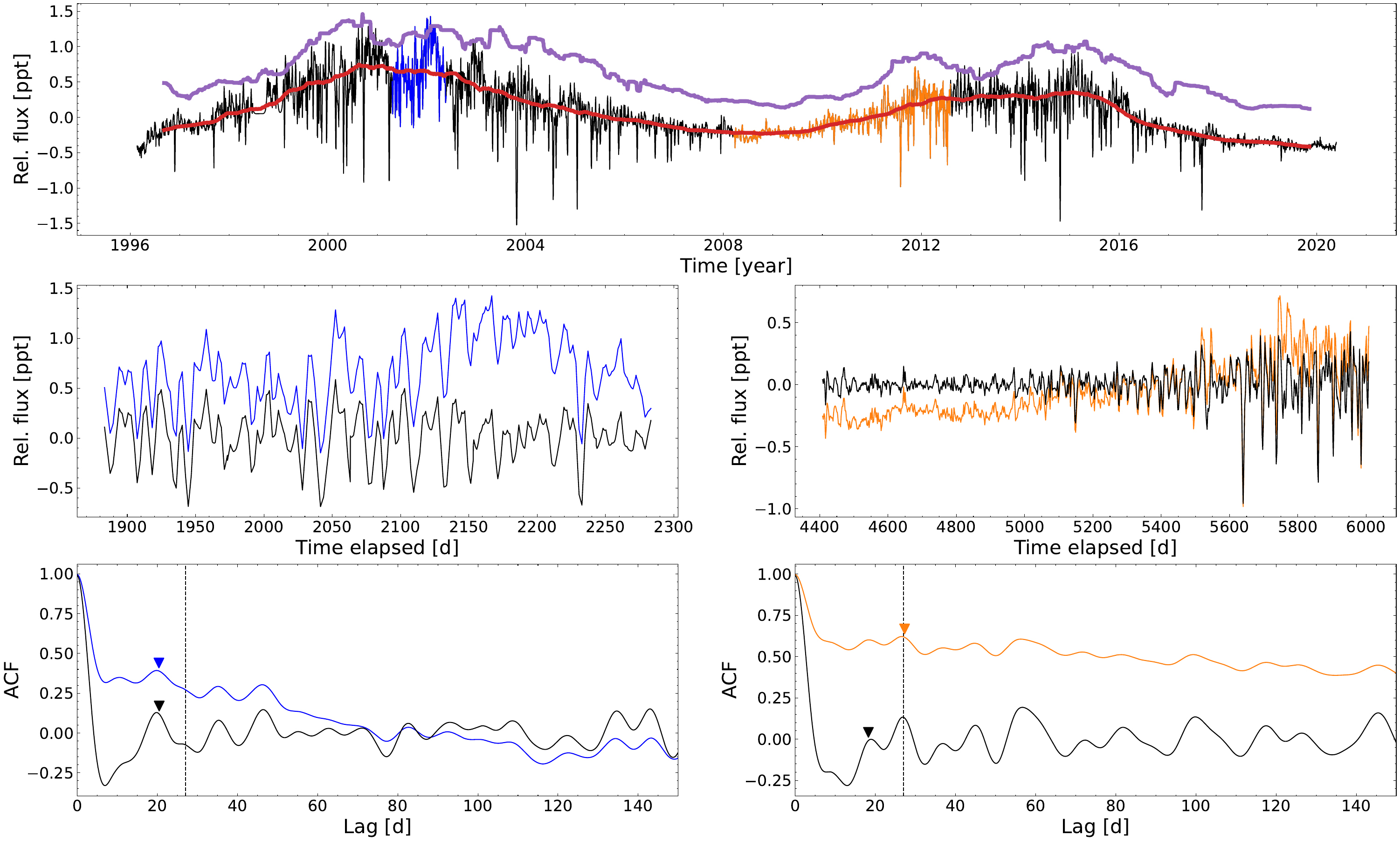}{1.0\textwidth}{}}
\caption{Upper panel: TSI light curve from VIRGO (VTSI; black). The brightness (\textbf{red}) and $R_\mathrm{var}$ (purple) over time computed from a sliding 400-day window are plotted as well (see text for explanation). Middle panels: examples of two segments of length 400 days (blue) and 1600 days (orange). Each original segment is plotted in its respective color, while the \textit{Kepler}ized version is plotted in black. Lower panels: autocorrelation functions are plotted using the same colors, with a colored marker at the location of the peak chosen as the correct period by ACF-MMA. The black dotted vertical line is the actual solar rotational period of 27 days.}
\label{fig:tsi}
\end{figure*}

We are primarily interested in the ability of ACF methods to determine rotational periods from \textbf{Sun-like} light curves. We do not test Lomb-Scargle periodograms because they are known to be less well-suited for analyze \textbf{\textit{Kepler} main sequence} light curves than the ACF \citep{McQuillan2013, Aigrain2015}. To search for rotational periods in our light curves, we employ an ACF method similar to that of \citet{MMA2014}, which we call ACF-MMA. Despite new advances in period detection such as the GPS method, we choose to base our timeseries analyses on ACF-MMA because of its established robustness in \textbf{stellar rotational period detection} as well as its familiarity and usage among the stellar astrophysics community. \textbf{The majority of the periods detected by ACF-MMA are from main sequence stars that have more periodic light curves than the typical Sun-like star, however.} We tested our version against a set of stars analysed by \citet{MMA2014} to confirm it returns very similar results. The manner in which our version of ACF-MMA differs from the original is that it tries different smoothing windows based on the observing cadence of the time series, since our data come from several different sources. 

\textbf{Other variations of ACF period detection methods have been employed by other authors. One approach, which we call ACF-TP (tallest peak), selects the peak with the largest local peak height (LPH: the difference between each peak in the ACF and the average of its adjacent troughs) within a chosen interval, such as 1-70 days. Another approach is the ``cleaned ACF'' (ACH-CLN) method described in Section 2.3 of \citet{Basri2022}. In addition, a ``goodness criterion" that ignores ACF peaks with LPH below some threshold has been used (e.g., \citep{MMA2014}, \citet{Reinhold2021}). We briefly compare ACF-MMA with these methods later on.} Our purpose here, however, is not to find a better method of using the ACF to find periods, but rather to test the role of faculae in period determinations. 

\subsection{Solar Datasets} \label{sec:data}
\begin{figure*}[htb!]
\gridline{\fig{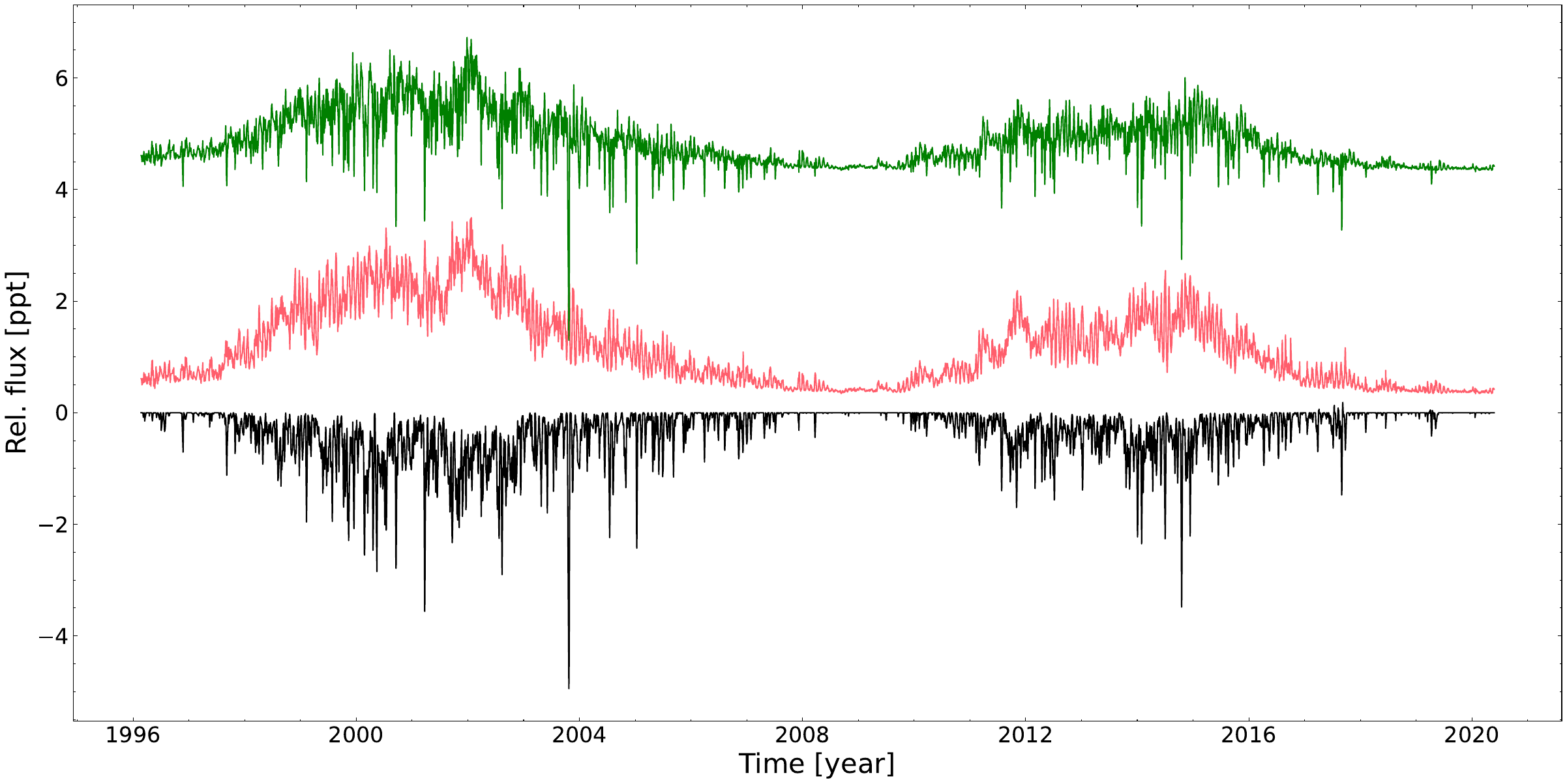}{1.0\textwidth}{}}
\caption{SATIRE-S reconstruction of the TSI (green, offset) with the individual facular (red) and spot (black) components also shown. The light curve spans the same dates as the VTSI in \figr{tsi}.}
\label{fig:satire_cycle}
\end{figure*}

We utilize two forms of empirical solar data and two forms of modeled light curves to conduct our experiments. The TSI has been measured from space for several decades \citep{SATIRE}. We use a version of these data \textbf{(available online)} from the VIRGO \textbf{(Variability of Solar Irradiance and Gravity Oscillations)}/PMO6-V instrument on SOHO (\textbf{Solar and Heliospheric Observatory}) for Cycles 23 and 24, and refer to this dataset as ``VTSI" \textbf{(1996–2020, cadence=1hr)}. Note that although the upper panel of (\figr{tsi}) shows it as relative flux, the absolute flux levels have been preserved and the scale simply shifted so that the median absolute flux is zero. Another version of this data that has been analysed for facular and spot components separately is also available \textbf{from the Spectral And Total Irradiance REconstructions for the Satellite era (SATIRE-S)}. We refer to \textbf{this dataset} as ``SATIRE-S" \textbf{(1996–2020, cadence=1d)}. \textbf{This} daily reconstruction of the TSI (\figr{satire_cycle}) is based on full-disc intensity and magnetogram images of the Sun \citep{SATIRE}. It is able to accurately reproduce TSI variations on timescales of a day or more by modeling the effects of solar magnetic activity into a separate spot and facular signal. \citet{SATIRE} have also modelled the Sun's solar spectral irradiance (SSI), providing the flux per unit wavelength.

\begin{figure*}[htb!]
\gridline{\fig{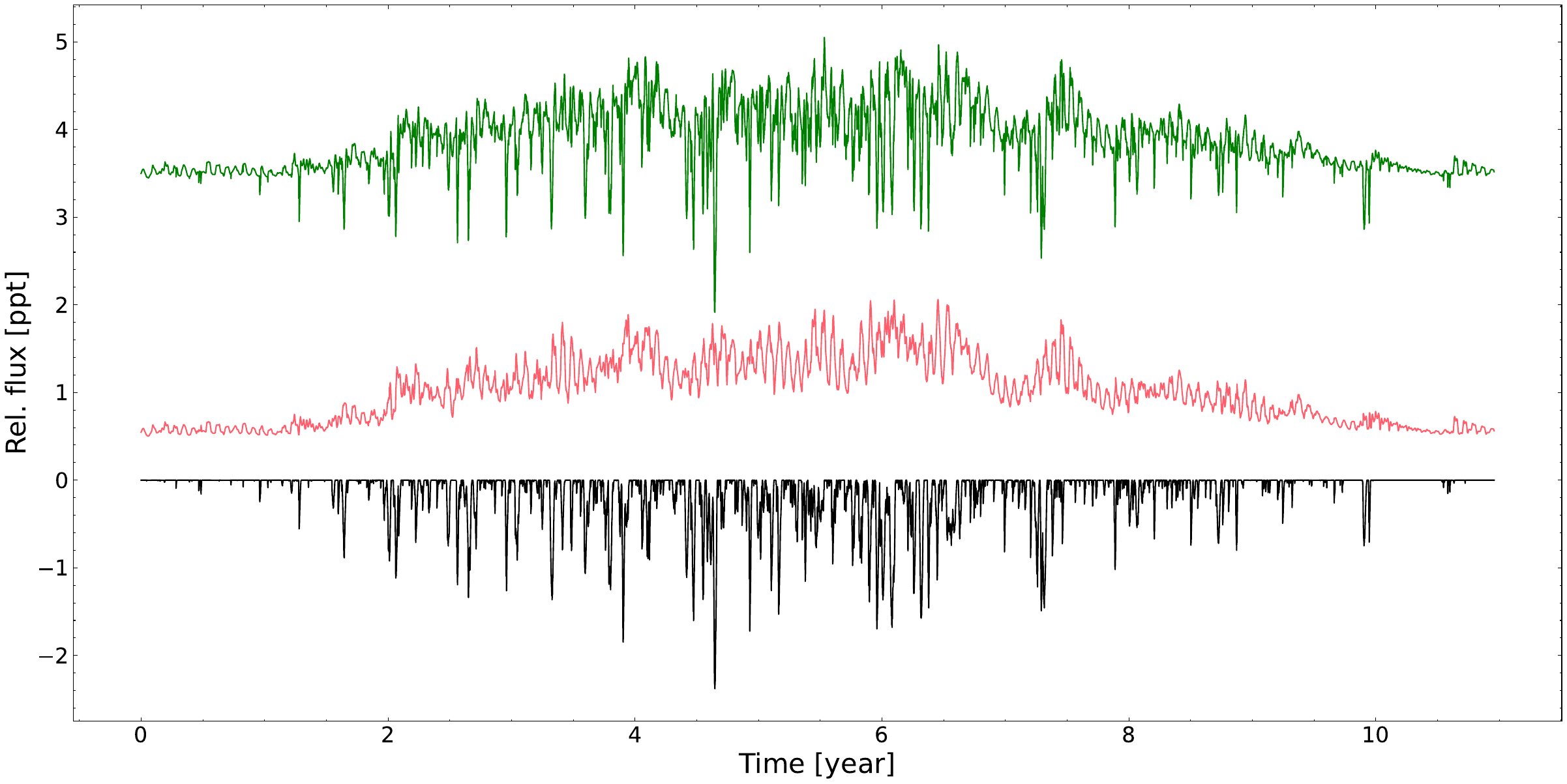}{1.0\textwidth}{}}
\caption{Cycle 4 out of the 25 modeled MPI solar cycles (green, offset) with the individual facular (red) and spot (black) components also shown. Note that in our analysis we stitch all 25 cycles as one contiguous light curve.}
\label{fig:mpi_cycle}
\end{figure*}

\begin{figure*}[htb!]
\gridline{\fig{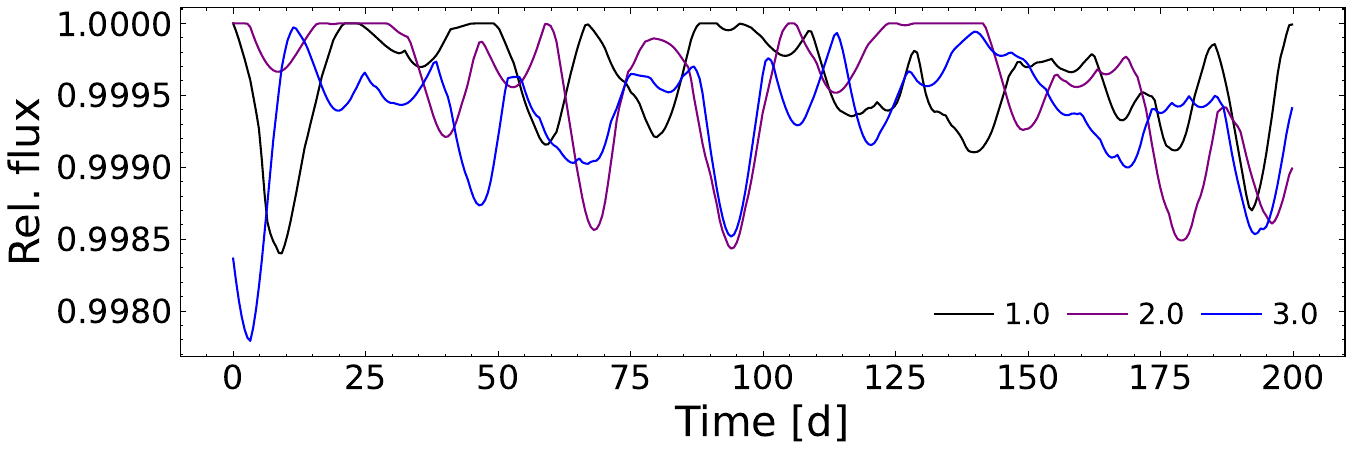}{0.7\textwidth}{}}
\caption{Examples of 200-day segments of light curves with fixed spot lifetime from our analytic spot model (the average spot number and size are also fixed). Lifetimes of 1.0 (black), 2.0 (blue), and 3.0 (purple) rotation periods are shown (they have not been \textit{Kepler}ized).}
\label{fig:analytic_model}
\end{figure*}

To extend our analysis to a larger set of data, we also utilize models from the {\it Max Planck Institute for Solar System Research} group \citep{Reinhold2021} which we refer to as ``MPI" \textbf{(1700–2010, cadence=0.25d)}. \textbf{Their paper discusses effects of various metallicities and inclinations on a metric similar to our bulk accuracy defined below}, and the light curves were kindly supplied to us by those authors. These models are constructed \textbf{at various inclinations} from photometric models of solar activity informed by data from the last 25 solar cycles, then translated to the \textit{Kepler} passband. \textbf{During our analyses, we use all 25 solar cycles with solar metallicity stitched together into one contiguous light curve}. An example of one model cycle \textbf{at solar inclination} is shown in \figr{mpi_cycle} with its respective spot and facular components. Finally we generated simple \textbf{300-rotation} analytic spot-only models (\figr{analytic_model}) using the methods of \citet{BasriShah2020} to further study the contribution of starspot lifetimes to the \textbf{bulk ACF accuracy}. \textbf{\tabr{datasets} summarizes these datasets.}

\begin{table*}[htb!]
\begin{tabular}{|l|l|l|l|l|}
\hline
\textbf{Dataset} & \textbf{Timespan}  & \textbf{Cadence} & \textbf{Passband}   & \textbf{Reference} \\ \hline
VIRGO/PMO6-V TSI (VTSI)    & 1996–2020 & 1 hr    & Bolometric                     &   \citet{Finsterle2021}        \\ \hline
SATIRE-S TSI        & 1996–2020 & 1 d     & Bolometric                     &   \citet{SATIRE}        \\ \hline
SATIRE-S Integrated SSI        & 1996–2020 & 1 d     & Red, Optical, \textit{Kepler}, Blue, UV &  \citet{SATIRE}         \\ \hline
MPI Model           & 1700–2010 & 0.25 d  & \textit{Kepler}                &   \citet{Reinhold2021}        \\ \hline
Analytic Spot Model & 300 rot  & 1/50 rot & Bolometric                     &   \citet{BasriShah2020}        \\ \hline
\end{tabular}
\caption{\textbf{Summary of light curve datasets used throughout the paper, described in \secr{data}.}}
\label{tab:datasets}
\end{table*}

\subsection{Monte Carlo Segment Choice and \textbf{Bulk} ACF Accuracy}

Our goal is to understand under what conditions observations on \textbf{Sun-like stars} yield the most accurate ACF period measurements. The light curves described above can be sampled in a variety of ways by breaking them up into observing windows (``segments"), sometimes chosen based on selected portions of an activity cycle, rebinning them, renormalizing them, or other operations. In our analysis, segments from the light curves are chosen independently and thus are allowed to overlap. The middle panels of \figr{tsi} show examples of two segments of different length randomly chosen from the entirety of the light curve in the upper panel.

\begin{figure*}[htb!]
\gridline{\fig{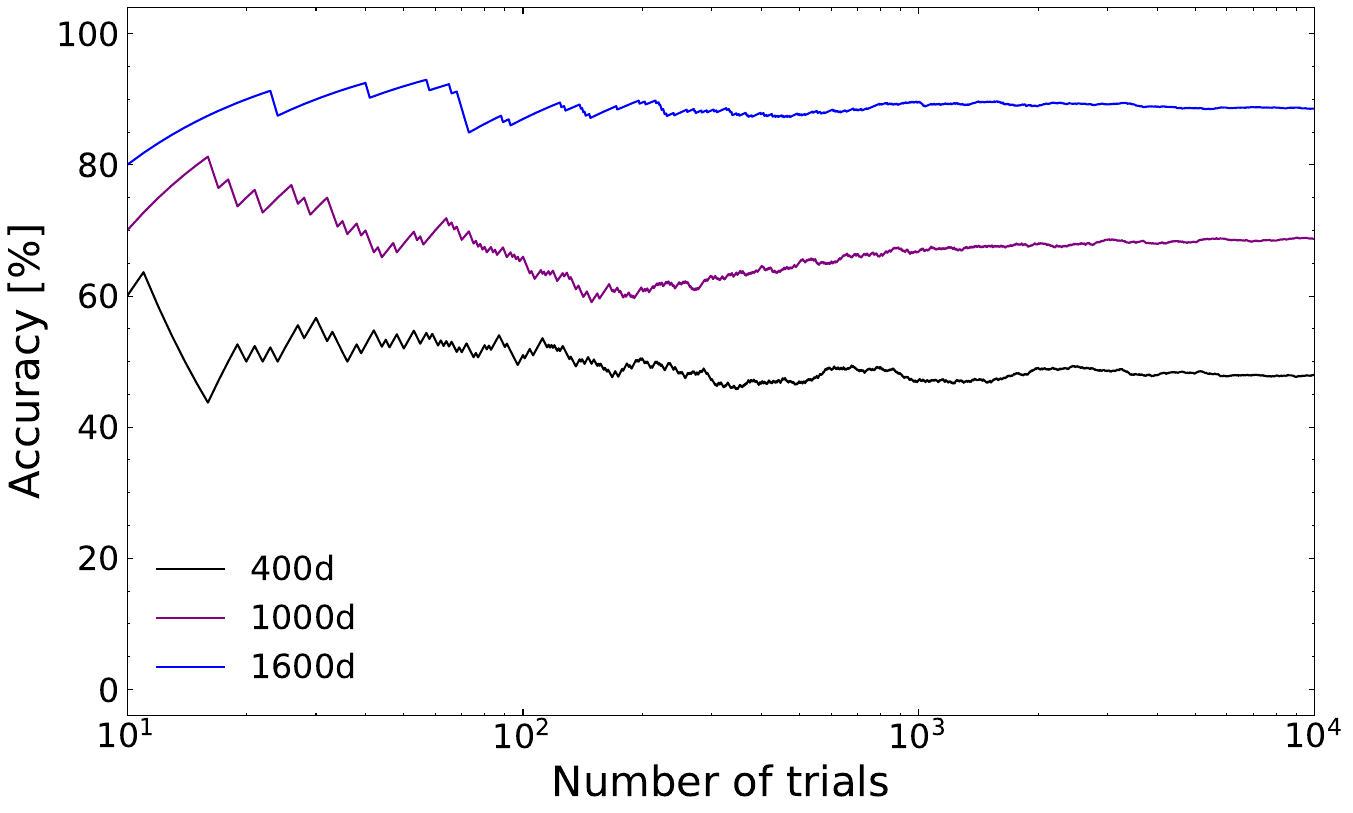}{0.6\textwidth}{}}
\caption{\textbf{Bulk} ACF accuracy \textbf{on the VTSI} as a function of Monte Carlo sample size \textbf{for segments of size 400d (black), 1000d (purple), and 1600d (blue)}. Convergence can be seen at sample sizes approaching 10,000. Note that the $x$-scale is logarithmic.}
\label{fig:convergence}
\end{figure*}

We employ a Monte Carlo approach to sample 10,000 randomly located segments of a fixed duration from the entire light curve, running our ACF analysis on each segment. \textbf{We define the bulk accuracy for any test case as the percentage of segments whose individual ACF determinations were within 10\% of the correct period (taken to be 27 days). We chose this tolerance level partly to cover the range of possible periods induced by solar-like differential rotation}. This \textbf{sample size} was found to be a good compromise between computation time and convergence of the bulk accuracy. \textbf{As an example,} \figr{convergence} shows the convergence of the \textbf{bulk accuracy} for three Monte Carlo samples \textbf{on the VTSI} where the segment sizes were short (\textbf{400d}), medium (1000d) and long (\textbf{1600d}). The bulk accuracy levels off near sample sizes of order of $10^4$, and similar results were found for changes in the various other variables we analyze. \textbf{For each of our light curve datasets, we perform Monte Carlo experiments for segment sizes ranging from 100–2000d.} 

In addition to looking at the proportion of nearly correct (accurate) periods for a given case, we examine the dispersion of the inferred periods about the \textbf{correct value. The distribution of inferred periods is not a normal distribution, for example sometimes showing enhancements at one or more harmonics of the period, as well as being skewed to longer rather than shorter periods. What we mean by a smaller dispersion is that there are fewer points well away from the correct period. We found that the dispersion of measured periods generally grows smaller as the bulk accuracy increases. We did not find it useful to delve further into the statistics of the dispersion.}

There are two ways in which we characterize the activity level of a given segment. For the first we adopt the ``range'' ($R_\mathrm{var}$) metric for stellar variability as defined by \citet{Basri2010}. The range is computed as the difference between the $95^\mathrm{th}$ and $5^\mathrm{th}$ percentile brightness of the differential light curve. In the context of \textit{Kepler} data, using this definition suppresses the impact of transients such as flares and transits on the photometry. $R_\text{var}$ is known to be correlated with magnetic activity, as the amplitude of flux variations increases with brighter faculae and darker spots. It has been fairly widely utilized by other authors, and produces similar results to other photometric activity measures that have been proposed and used.

The second method of quantifying the activity level during a segment involves defining the absolute brightness of the segment to be the median flux of the segment relative to the entirety of the (unnormalized) light curve. This is motivated by the fact that the absolute flux of the faculae-dominated Sun increases as it becomes more magnetically active. We found that $R_\text{var}$ and the brightness give similar results in quantifying activity level for \textbf{Sun-like} light curves. In our results we discuss both metrics. The upper panel of \figr{tsi} shows the brightness (red) and $R_\text{var}$ (purple) over time obtained by and sliding a 400-day window across the entire VTSI light curve\textbf{, starting from $t=0\text{d}$ to $t=400\text{d}$ and advancing by one cadence each time}. \textbf{The choice of 400 days here is pedagogical and somewhat arbitrary; in general, we choose 400-day segments throughout the paper as discussion examples because that length resembles a year and contains enough rotation periods to allow a period determination.}

\subsection{\textit{Kepler} systematics and ``\textit{Kepler}ization''}

Although \textit{Kepler} data is well-suited for exoplanet science, namely analyzing the relatively short dips of transits and eclipses, using it for identifying and characterizing stellar magnetic activity of \textbf{Sun-like} stars is a more difficult task. This is not only due to the fact that \textbf{Sun-like} stars are inherently difficult candidates for rotational period detection due to their often aperiodic spot signatures, but also due to systematic effects caused by the \textit{Kepler} data reduction pipeline. In addition to removing instrumental trends, the PDC-MAP pipeline \citep{Smith2020} unfortunately can remove or redistribute the power of long-term trends on the order of more than 20 days \citep{Aigrain2015}. In practice, any periodicities above 30 days can be expected to have some component of an instrumental trend, while periodicities over 50 days should be interpreted with suspicion \citep{Santos2019}. In our analyses of light curves, we implement the oft-employed procedure called ``\textit{Kepler}ization'' (\citealt{Basri2018, Reinhold2021}) by subtracting quadratic trends fitted to separate 90-day ``quarters" in the light curve. We do this primarily to simulate what happens to the absolute brightness (which our TSI light curves have) when observed by an instrument without absolute calibration. In a \textit{Kepler}ized light curve there may exist jumps in flux at the boundaries of the quarters, which is similar to what occurs in the actual \textit{Kepler} data (after each quarter the spacecraft was rotated). 

This allow us to answer a question specific to \textit{Kepler} data: whether the photometry being differential (\textit{Kepler}ized) or absolute has a significant effect on the ACF accuracy. For each set of 10,000 randomly chosen segments from a full light curve, we determine the \textbf{bulk accuracy} when all of the segments are each either \textit{Kepler}ized or left in their original absolute brightness form. The lower panels of \figr{tsi} show two examples of original/\textit{Kepler}ized segments and their ACFs. We found that in general \textit{Kepler}ization has the effect of making most of the ACF local peak heights  larger. This does not always improve the accuracy of the period determination however, because the ACF-MMA method examines only the first two peaks, and the heights of spurious peaks can sometimes become more dominant. This occurs in the bottom right panel of \figr{tsi}, where because the highest of the first two peaks is chosen for the period, the enhancement of a small peak near 10 days causes the peak near 20 days to be chosen even though the (third) peak at the correct period is even higher. 

\subsection{Other observational factors}
Next, we examine how the spectral passband affects the \textbf{bulk accuracy}. TSI light curves follow the \textbf{Sun's} bolometric luminosity, but stellar photometric monitoring cannot observe that directly. Using the SATIRE-S \textit{spectral} irradiance curve, we run our Monte Carlo approach on \textbf{solar} light curves in different restricted passbands (with the same start and end dates as the VTSI). Specifically, we analyze optical (400–700 nm), red (620–750 nm), blue (380–500nm), and \textbf{ultraviolet} \textbf{(UV; }10–400 nm) passbands, along with the Kepler passband. We treat all passbands except \textit{Kepler} as a boxcar window over the specified wavelength range. The facular signal should be enhanced at shorter wavelengths since faculae are hotter than the quiet Sun. Additionally, they are also broadly associated with active regions, which are much brighter in the UV. Because of their much lower temperatures, the signal due to sunspots should be attenuated at shorter wavelengths. Thus we expect the \textbf{bulk} accuracy to be higher on light curves with shorter wavelength passbands since faculae are more periodic than spots.

\textit{Kepler}'s remarkably low noise level was achieved to allow the detection of transits by Earth-sized planets but it also allowed the detection of photometric variations as low as those produced by the Sun. Solar experiments achieve even lower noise levels. To understand the effects of \textbf{Gaussian noise} in general (not specifically Poisson or instrumental noise), we rerun our Monte Carlo analysis on \textbf{the SATIRE-S net light curve} \textbf{multiple times}, \textbf{each time with a certain amount of} \textbf{Gaussian noise} \textbf{injected (0-500 ppm)}. Noise levels for the fainter \textit{Kepler} targets are higher than this. For all other analyses, we do not add any extra noise, ensuring that the effect of the independent variable of interest on \textbf{bulk accuracy} was not confounded with noise.

\subsection{\textbf{Astrophysical effects}}

\textbf{There are two quantities that are inherently random when observing a star for a fixed window of time. One of these is the star's inclination, $i$. Assuming the distribution of all stellar inclinations to be isotropic, the distribution of $i$ is proportional to $\sin i$, meaning that inclinations closer to 90° (edge-on) are more likely than 0° (spots do not rotate out of view), with the mean inclination being about 57°. Using the MPI models, we ran our Monte Carlo analysis on the 25 solar cycles simulated at different viewing inclinations.}

\textbf{The second of these factors is the star's activity level. At solar maximum spots dominate the \textbf{TSI} variations \textbf{on rotational} timescales (although faculae dominate the absolute brightness), yielding an overall less periodic signal}. But at solar minimum, the faculae dominate and tend to produce more periodic light curve segments relative to the \textbf{solar maximum}. If one observes a star for 400 days \textbf{at a given time}, its phase in the stellar magnetic cycle is unknown and can be regarded as essentially random. Our Monte Carlo approach \textbf{marginalizes over this} by taking randomly located segments throughout the light curve to imitate various possible observing windows of a specific length. \textbf{In order to isolate the effects of activity level, we record the activity (absolute brightness and $R_\text{var}$) of each of the 10,000 segments in our Monte Carlo analyses for the SATIRE-S and MPI light curves, allowing us to analyze the bulk accuracy as a function of activity. }
 
The relatively diffuse spatial distribution and longer lifetime of facular regions on the solar surface allows them to more easily generate periodic light curves (\citealt{Chapman1997, Basri2018}). On the other hand spot groups, which in the solar case rarely last longer than one rotation period, can evolve substantially as they rotate across the disc, interfering with the \textbf{periodic} rotation signal. In order to understand the separate effects of spots and faculae, we leverage the availability of individual spot and facular brightness variations for our analysis, something we could not do with just the VTSI.  We rerun our Monte Carlo analysis on the SATIRE-S net light curve multiple times, each time scaling the facular signal up by fixed amounts varying from 1.5X to 4X and then reconstituting a net signal using the original spot signal. To extend our facular scaling analysis to a larger set of data, we ran an identical analysis on 25 contiguous \textbf{Sun-like} MPI cycles from 1700–2010.

One shortcoming of using the SATIRE-S light curves is that they are modeled on actual solar data interpreted by empirical models for light curves of faculae and spot, and not on less directly observable quantities such as spot lifetimes. We therefore are unable to study the effect of spot lifetimes with the SATIRE-S or MPI models. \citet{Nemec2022} show that ``nesting", a solar term referring to the case where new active regions appear in similar locations as recent former ones, causes the net signal (dominated by spots) to appear more periodic. It is, however, difficult to translate the quantitative effects of the nesting factor to specific spot lifetimes (which also depend on spot size in those more physical calculations). 

\citet{BasriShah2020} showed that pure starspot light curves are more periodic when starspots live for several rotations. To investigate how increasing spot lifetime improves the \textbf{bulk accuracy}, we analyze light curves produced utilizing the techniques from that paper. The models have spots with a fixed rotational lifetime and maximum size. \textbf{We employed 6-spot models (in their terminology) to generate 300 27-day rotation light curves with spot lifetimes of 1.0, 1.5, 1.75, 2.0, 2.5, and 3.0 rotation periods (keeping the spot maximum size fixed).} \figr{analytic_model} shows an example of 6-spot light curves at spot lifetimes of 1.0, 2.0, and 3.0 rotational periods. Repeating our Monte Carlo analysis with these spot light curves, we computed the \textbf{bulk accuracy} at the various spot lifetimes. We did not combine these models with a facular signal; it is apparent that having a facular component generally improves the \textbf{bulk accuracy} so our values for the accuracy serve as a lower bound. These models are clearly a departure from the more physical and \textbf{Sun-like} light curves used above. 

\section{Results and Discussion}
\subsection{\textbf{Comparison of ACF Period Determination Methods on the SATIRE-S Dataset}}
\label{sec:method_comparison}
\textbf{We first discuss the difference in our bulk accuracy results that arise from employing different ACF period determination methods. We compare our bulk accuracy results on the SATIRE-S dataset between ACF-MMA and three other strategies we tested: (1) ACF-TP, which selects the peak between 1–70d with the largest LPH; (2) ACF-MMA-LPH, which identical to ACF-MMA but only returns a period if the dominant peak has LPH $>0.1$; and (3) ACH-CLN \citep{Basri2022}. The SATIRE-S dataset contains the individual spot and facular contributions to the net light curve, so it is instructive to show the results for these separate components.}

\begin{figure*}[htb!]
\gridline{\fig{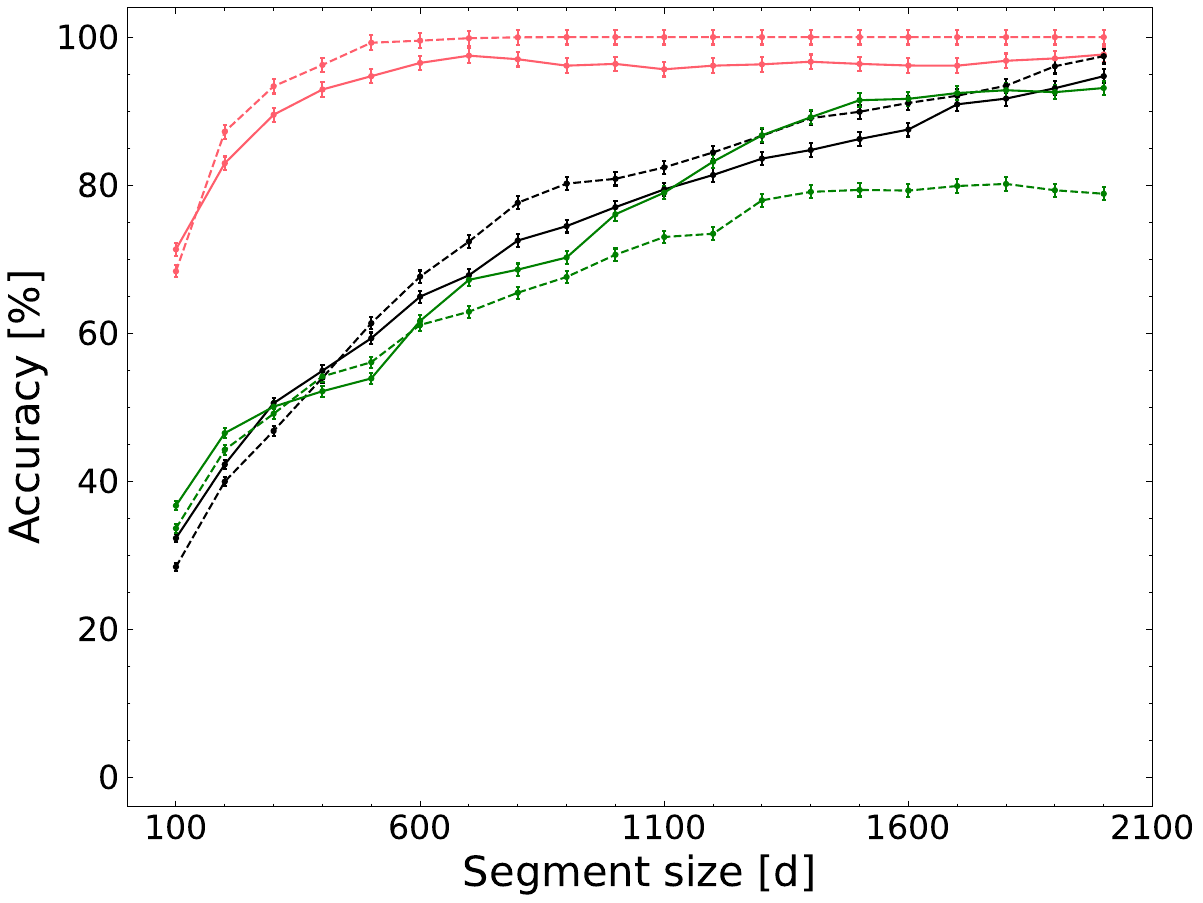}{0.5\textwidth}{(a) ACF-MMA}\fig{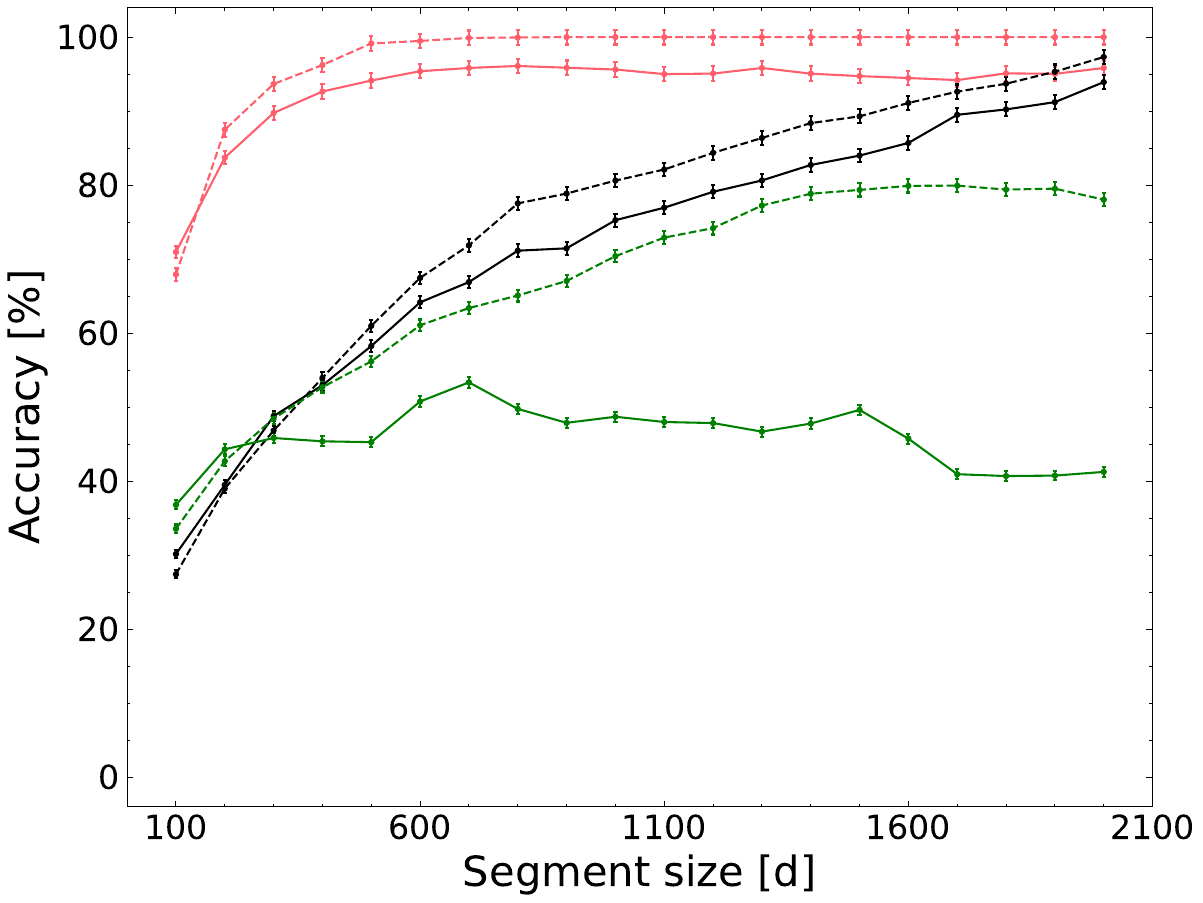}{0.5\textwidth}{(b) ACF-MMA-LPH}}
\gridline{\fig{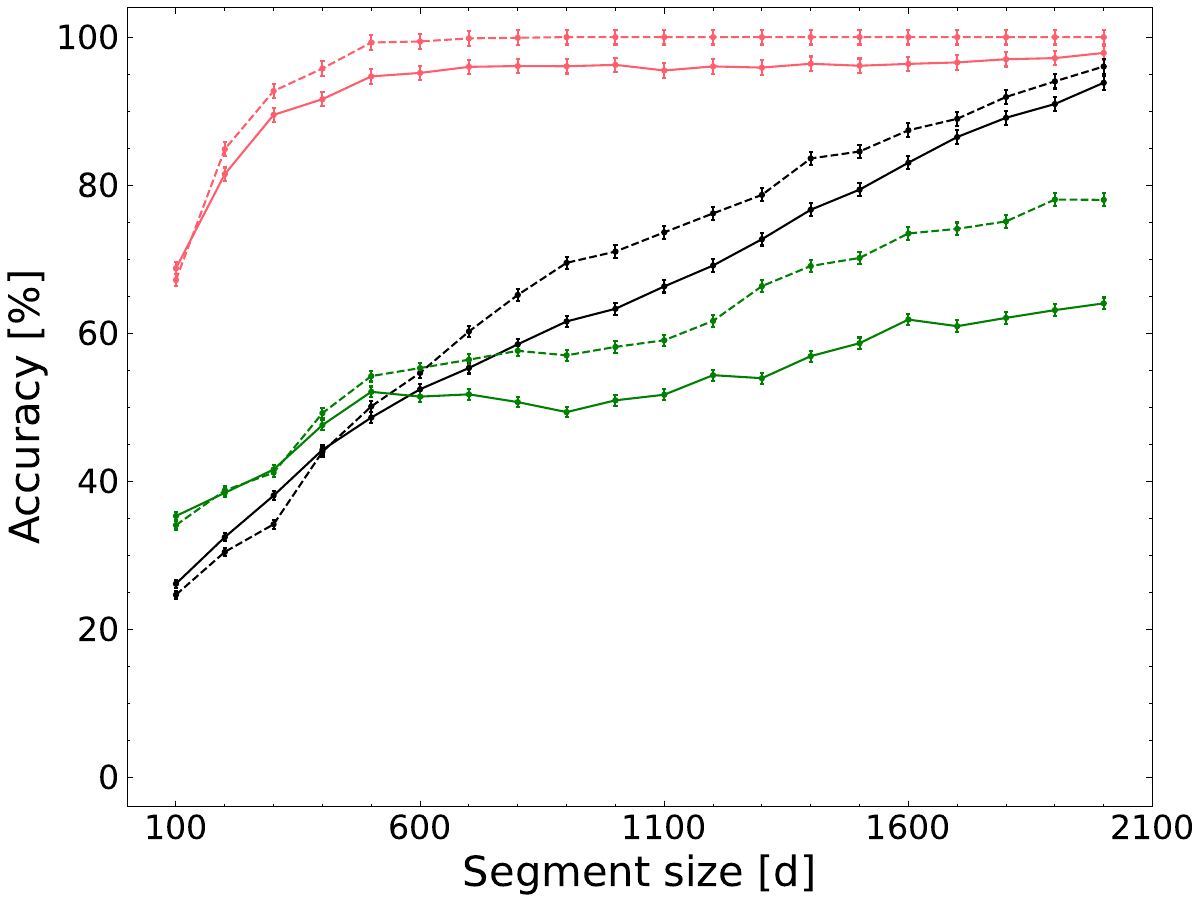}{0.5\textwidth}{(c) ACF-TP}\fig{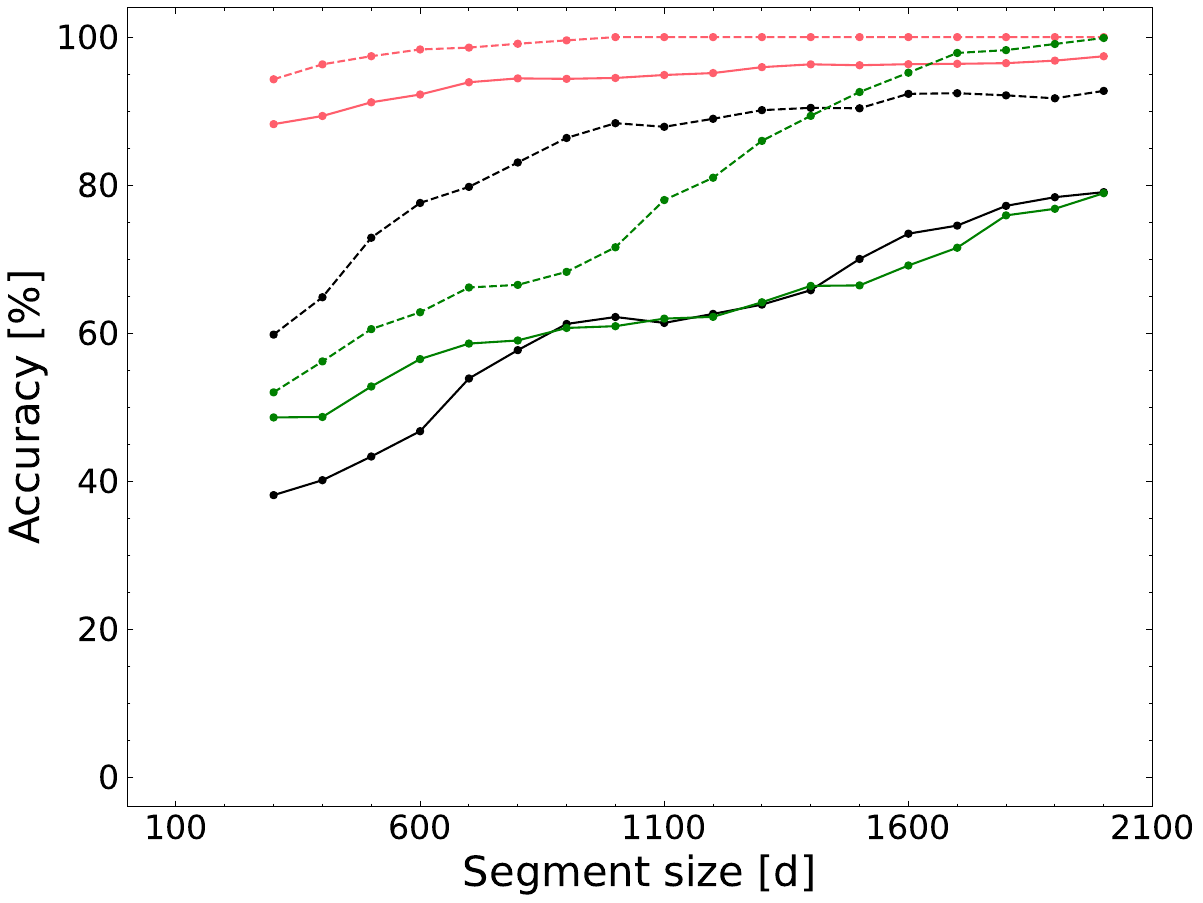}{0.5\textwidth}{(d) ACH-CLN}}
\caption{\textbf{Bulk ACF accuracy on the spot (black), facular (red), and net (green) signals from the SATIRE-S dataset, obtained with several different ACF period determination methods: ACF-MMA (top-left), ACF-MMA-LPH (top-right), ACF-TP (bottom-left), and ACH-CLN (bottom-right). See the text for descriptions of each. The error bars are the square root of the number of accurate measurements divided by the Monte Carlo sample size. The solid and dashed lines correspond to accuracies when the segments were un\textit{Kepler}ized and \textit{Kepler}ized, respectively.}}
\label{fig:method_comparison}
\end{figure*}

\textbf{\figr{method_comparison} shows the results obtained by running Monte Carlo analyses on the facular (red), spot (black), and net (green) signals with each of the four ACF period determination methods. Each panel is a prototypical example of how we display many results throughout the paper, presenting the relationships between bulk accuracy and segment size. \textbf{The error bars are similar to Poisson errors since they are the square root of the number of accurate measurements divided by the Monte Carlo sample size. From hereon, all error bars in the \textbf{bulk accuracy} refer to this definition. We note this does not include various possible systematic errors, such as noted in Section 3.7.} \textbf{In addition, the solid and dashed lines in the plots of bulk accuracy correspond to accuracies when the segments were un\textit{Kepler}ized and \textit{Kepler}ized, respectively.}
}

\textbf{It is immediately clear that the bulk accuracies for the three signals increase with segment size, although the exact bulk accuracies vary significantly between methods (up to a 20\% difference). The one exception is with ACF-MMA-LPH, where the un\textit{Kepler}ized bulk accuracies show a decrease with increasing segment size. This is because the longer un\textit{Kepler}ized segments feature the activity cycle variability, which adds a trend to the ACF and decreases the LPH of each peak (the bulk accuracies of the spot and facular signals are impacted less since most of their dominant peaks have LPH $> 0.1$ even before \textit{Kepler}ization). Thus when employing goodness criteria on the local peak height, \textit{Kepler}ization, or removing these long term trends, is helpful. }

\textbf{For the individual SATIRE-S facular (red) components, it is clear that the they are quite accurate even for the shortest segments, and reach full \textbf{bulk accuracy} for segments longer than about 500 days. The spot (black) components, on the other hand, remain below 50\% \textbf{bulk accuracy} for segments shorter than 500 days and only approach the \textbf{bulk accuracy} of the facular measurements for segments nearly 2000 days long (more than 5 years). \textbf{The bulk accuracies of the net signal tend to follow those of the spots, but sometimes the spot signal is even more accurate, despite the net having the facular contribution.}}

\textbf{Since our goal is to study faculae rather than ACF methods, the rest of the figures only feature results from ACF-MMA. We note that our general qualitative conclusions, such as the \textit{Kepler}ized bulk accuracy increasing with segment size, follow regardless of the period determination method employed.}


\subsection{\textbf{Spectral Passband}}
\begin{figure*}[htb!]
\gridline{\fig{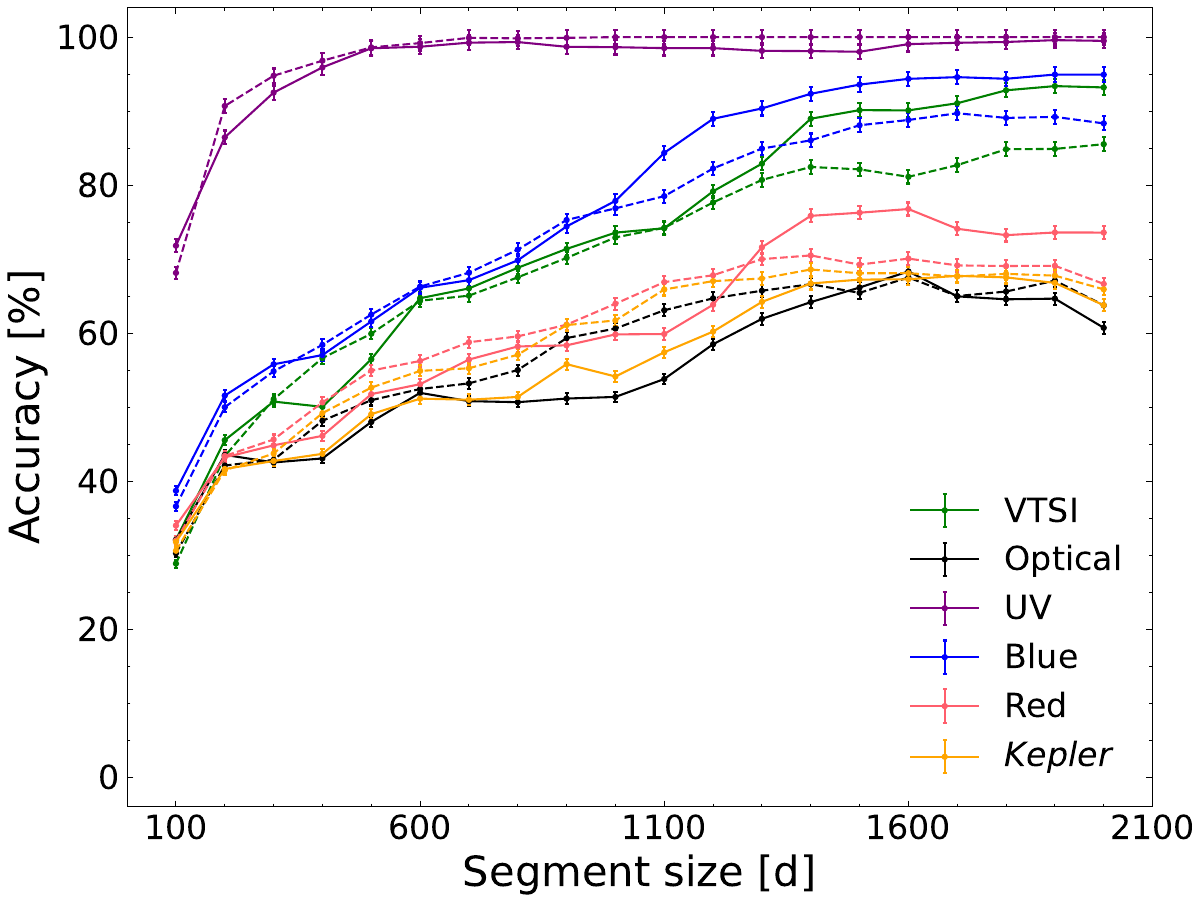}{0.6\textwidth}{}}
\caption{\textbf{Bulk ACF accuracy} for the VTSI (green), and for \textbf{SATIRE-S} solar light curves from the same dates obtained through different passbands. The passbands include optical (black), red (red), blue (blue), UV (purple), and \textit{Kepler} (orange). See the text for the specific wavelength intervals used.}
\label{fig:passband}
\end{figure*}

We ran our Monte Carlo analysis on the integrated solar output in \textbf{numerous} passbands, making use of the SATIRE-S SSI as described above. Our results are shown in \figr{passband}. \textbf{We include results from the VTSI light curve for comparison.} The least to most accurate passband period measurements (in that order) are optical, \textit{Kepler}, red, \textbf{VTSI}, blue, and UV. 

Somewhat unexpectedly the optical passband is the least accurate, even though it presumably contains more of a facular signal than the red passband. The \textit{Kepler} accuracies are quite close to those from the optical passband, and significantly worse than those from the \textbf{VTSI}. The red accuracies are significantly higher for longer segment lengths. TESS has a somewhat redder passband than \textit{Kepler}, but unfortunately its higher noise levels and much shorter observational windows make it very difficult to detect periods of Sun-like stars with that instrument. The VTSI signal apparently has a similar amount of periodicity to that of the blue signal. This might be because it contains very periodic UV information, even though that is only a tiny fraction of the total signal. The UV signal is driven by high contrast plage in the chromosphere, not the faculae \textbf{, and has very high bulk accuracy}. Overall, a very substantial increase in \textbf{bulk accuracy} can be obtained by observing at shorter wavelengths or including some UV signal, which is almost perfectly periodic \textbf{with respect to ACF methods}.

\subsection{\textbf{Photometric Noise Level}}
\begin{figure*}[htb!]
\gridline{\fig{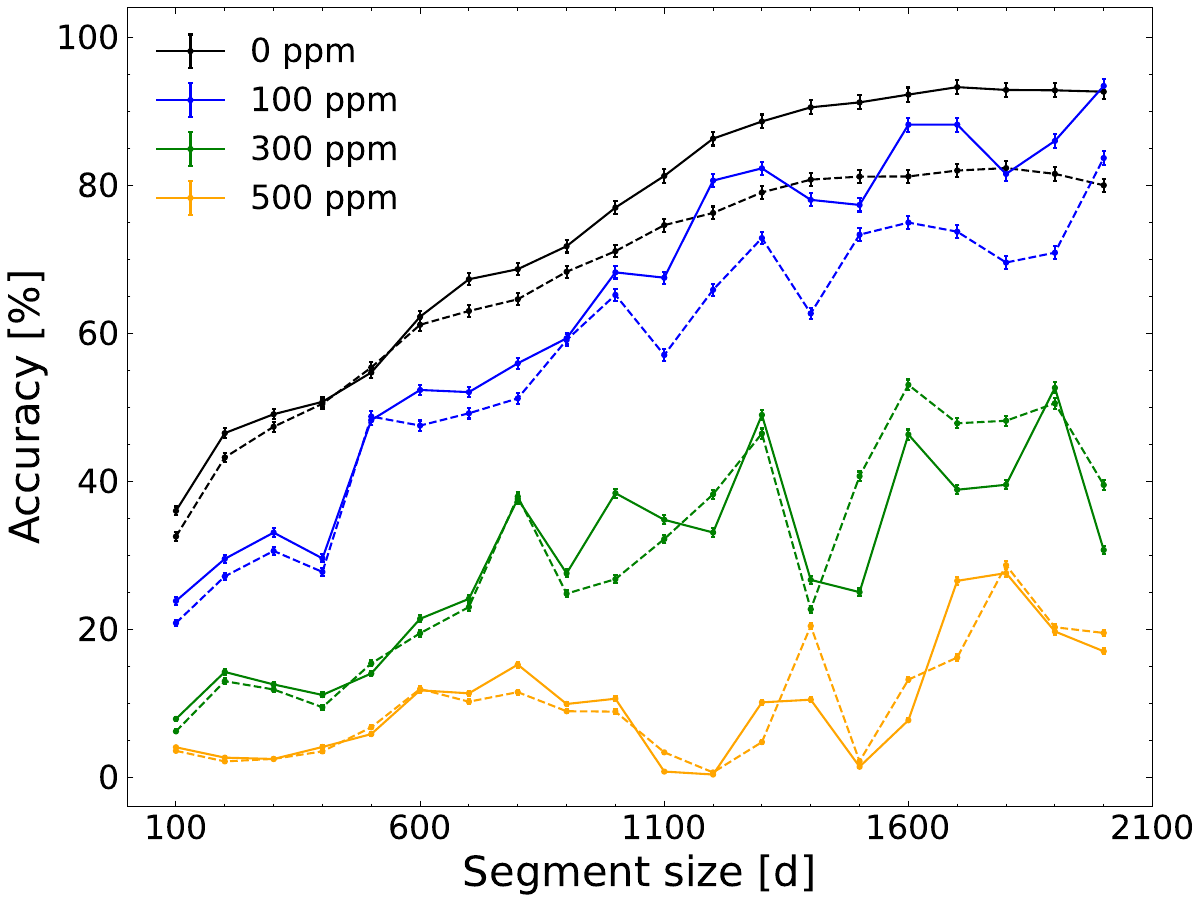}{0.6\textwidth}{}}
\caption{\textbf{Bulk ACF accuracy for the SATIRE-S net light curve with different amounts of \textbf{Gaussian noise} injected (see legend).}}
\label{fig:noise}
\end{figure*}

In \figr{noise}, we show the effect on the bulk accuracy of adding \textbf{Gaussian noise}. For all segment sizes, we see a negative correlation between \textbf{bulk accuracy} and added noise. Adding as little as 100 ppm of \textbf{Gaussian noise} sometimes decreases the \textbf{bulk accuracy} by over 10\%. \citet{Reinhold2021} conducted a similar noise analysis and found that when injecting Poisson noise to simulate observing fainter stars, the ACF local peak heights (relative to the adjacent dips) dropped. This can also affect which peak is chosen to represent the rotation period by ACF-MMA. 

\textbf{It is interesting that the bulk accuracies plotted against segment size for the noisy curves exhibit considerable scatter. We reran our analysis by reinjecting Gaussian noise and found the bulk accuracies again fluctuated randomly about a smooth trend. We saw similar behavior when using ACF-MMA-LPH, which only returns a period if its LPH is above 0.1. This suggests that our Poisson error bars are not very useful for light curves with significant noise, and that qualitative trends in the ACF can only be trusted at the 10\% level.}

\subsection{\textbf{Stellar Inclination}}
\begin{figure*}[htb!]
\gridline{\fig{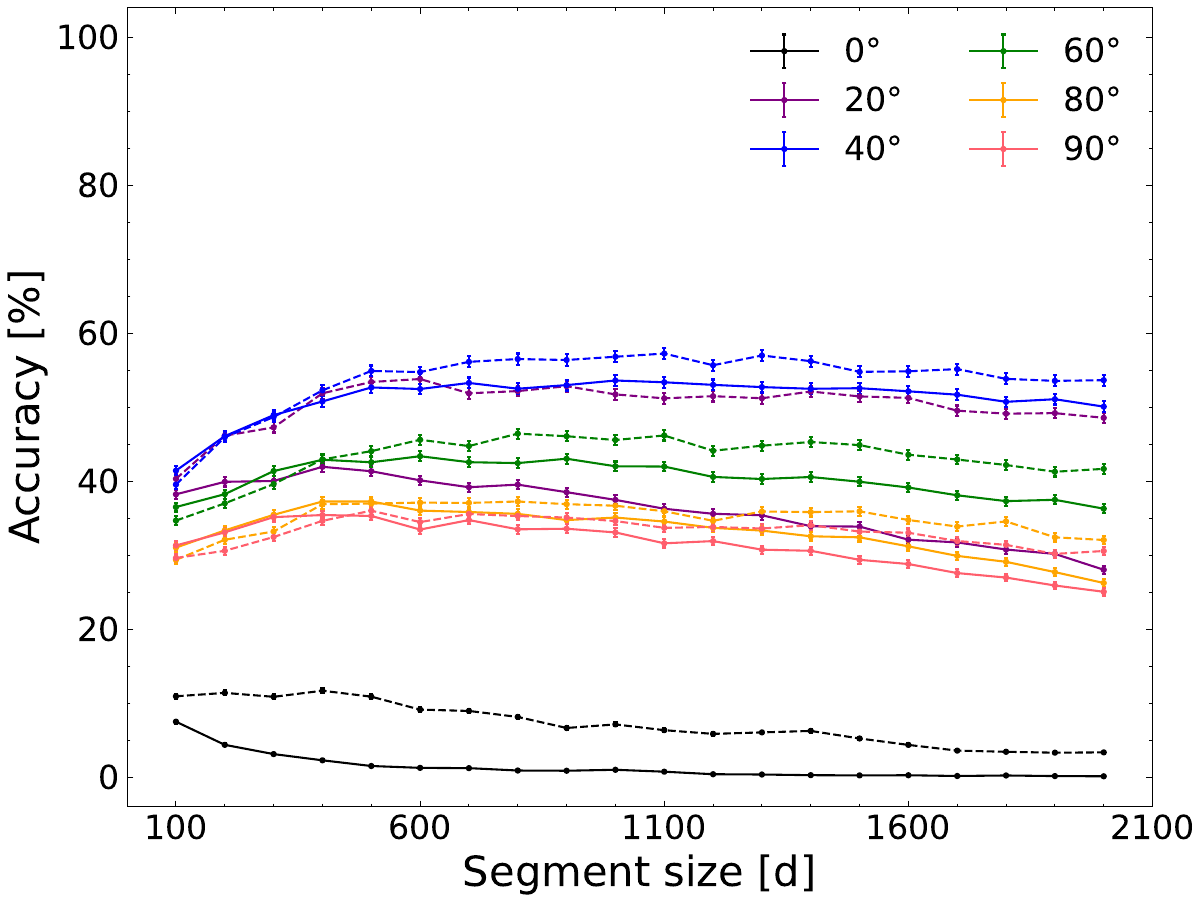}{0.6\textwidth}{}}
\caption{\textbf{Bulk ACF accuracy for the MPI net light curves simulated for different intermediate viewing inclinations (see legend).}}
\label{fig:mpi_inclination}
\end{figure*}

\textbf{\figr{mpi_inclination} shows our results for our Monte Carlo analyses on the MPI dataset recomputed at various viewing inclinations. The bulk accuracy is overall the largest for 40° inclination, consistent with the trends in  \citet{Reinhold2021} and \citet{Shapiro2016}. This is because faculae are brighter near the stellar limb, and if magnetic activity is confined to lower latitudes as in the Sun, it will be seen closer to the limb at all longitudes at moderate inclinations. As expected, for 0° inclination, the bulk accuracy is the lowest for all segment sizes since the flux variations are only due to active region evolution and not stellar rotation. For the rest of our results, the inclination is like that of the Sun, close to 90°.}

\textbf{Unlike the SATIRE-S dataset, the results using the MPI light curves do not show an increase in bulk accuracy with segment size. We discuss this discrepancy further in \secr{faculae_scaling}.}

\subsection{\textbf{Stellar Activity Level}}
\begin{figure*}[htb!]
\gridline{\fig{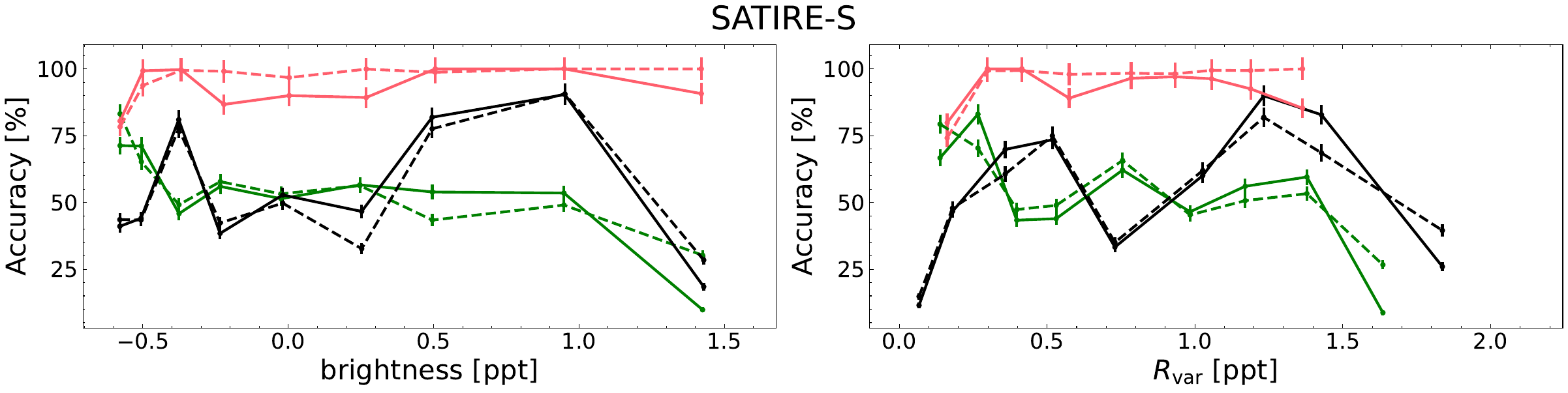}{1.0\textwidth}{}}\vspace{-20px}
\gridline{\fig{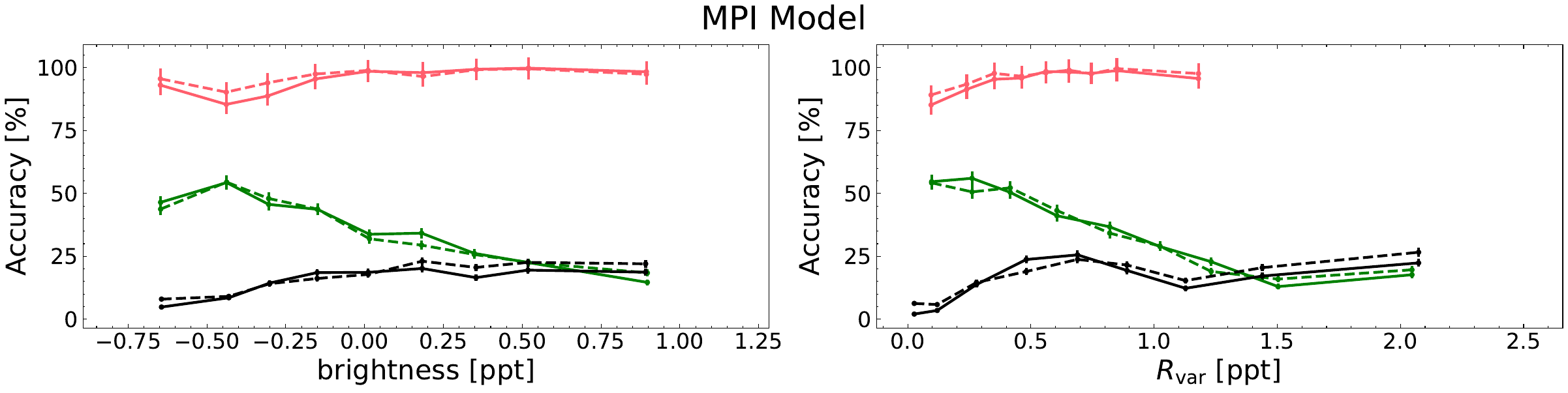}{1.0\textwidth}{}}
\caption{Bulk ACF accuracy as a function of brightness (left) and $R_\text{var}$ (right) for 10,000 400-day segments from SATIRE-S (top panels) and MPI (bottom panels). The red, black, and green colors correspond to the facular, spot, and net results, respectively. The solid and dashed lines correspond to accuracies when the segments were un\textit{Kepler}ized and \textit{Kepler}ized, respectively.}
\label{fig:accuracy_vs_activity_satire_and_mpi}
\end{figure*}

Another question of interest is to what extent the activity level of the star during an observational window matters. The segments used in the above trials are uniformly distributed in time across the entire solar light curve, so they do not test how the \textbf{bulk accuracy} is sensitive to activity levels. \textbf{Our results plotting the bulk accuracy versus activity level for 400-day segments taken the facular (red), spot (black), and net (green) SATIRE-S and MPI components are shown in \figr{accuracy_vs_activity_satire_and_mpi}. Each plot was generated in the following manner: the 10,000 periods in the Monte Carlo sample are binned by activity level (brightness/$R_\text{var}$) into nine bins of equal size. We chose to present the results from 400-day segments but the results are qualitatively similar for 100-day segments as well. Each point is located at the midpoint of its activity bin and represents the bulk accuracy of that bin. We note that the clarity of the behavior seen in \figr{accuracy_vs_activity_satire_and_mpi} varies based on the binning; our final choice shows the effects most clearly. Although not shown, we saw similar results with the VTSI compared to SATIRE-S.} 

Since we have knowledge of the facular signal, the brightness metric for all three components was derived from it directly (and solely). For both the SATIRE-S and MPI light curves, the facular signal remains at nearly full accuracy for all brightness or range values. This drives home the point that faculae by themselves provide a very periodic signal.

The behavior of the \textbf{bulk accuracy} from the SATIRE-S net light curve is highest at low activity levels and drops to a low value at the highest activity level in both metrics. The \textbf{bulk accuracy} from spots are higher at both small and large brightnesses of around -0.4 and 1.0 ppt. For the MPI models, the \textbf{bulk accuracy} from the spot signal increases slightly with brightness \textbf{while the net signal shows a clear decrease with brightness and $R_\text{var}$.} 

It is interesting that the \textbf{bulk accuracy} for the spot component in SATIRE-S is generally quite a bit higher than for the MPI cases. It approaches 100\% at brightness around 1.0 ppt, where the \textbf{bulk accuracy} of the net signal is actually lower than either of the individual \textbf{facular/spot} components. The cause might be a temporary active longitude, since the SATIRE-S curves are derived from the behavior of the actual Sun. In the case of $R_\text{var}$ for the SATIRE-S and net cases, the \textbf{bulk accuracies} peak at the lowest $R_\text{var}$ (activity minimum) as expected. The bulk accuracy of the MPI net signal shows a significant downward slope with increasing activity level, supporting the premise that at activity minimum, facular dominate the brightness variations and cause the net curve to be much more periodic, while the opposite occurs at activity maximum. \textbf{The results from SATIRE-S show a similar but less well-defined trend}.

\textbf{The results seen in \figr{accuracy_vs_activity_satire_and_mpi} are rather clear-cut for the MPI models, but the same cannot be said for SATIRE-S. In order to confirm that the difference is not caused by the fact that the SATIRE-S data only spans two cycles, we reran our analysis on just two out of the 25 contiguous MPI cycles, and obtained similar results to \figr{accuracy_vs_activity_satire_and_mpi}.}

It is evident from \figr{accuracy_vs_activity_satire_and_mpi} that the facular values of $R_\text{var}$ do not span the same  range as the spot and the net components. This is because the faculae generally induce weaker brightness contrasts than the spots, leading to a lower amplitude facular signal. Even though one might think that more active stars would have brighter faculae, \citet{Nemec2022} show that spot areas grow faster than facular areas, so the faculae will actually have less photometric effect on active stars. In any case, as we show below, if spots live for more rotations on active stars then spots will likely contribute more to the periodic signal than faculae. These longer rotational lifetimes are known to occur in active stars, aided by the fact that with shorter stellar rotation periods, spots that live for the same physical lifetime will have longer rotational lives on rapid rotators.

\subsection{How Bright do Faculae Have to Be to Induce Periodicity in the ACF?}
\label{sec:faculae_scaling}

\begin{figure*}[htb!]
\gridline{\fig{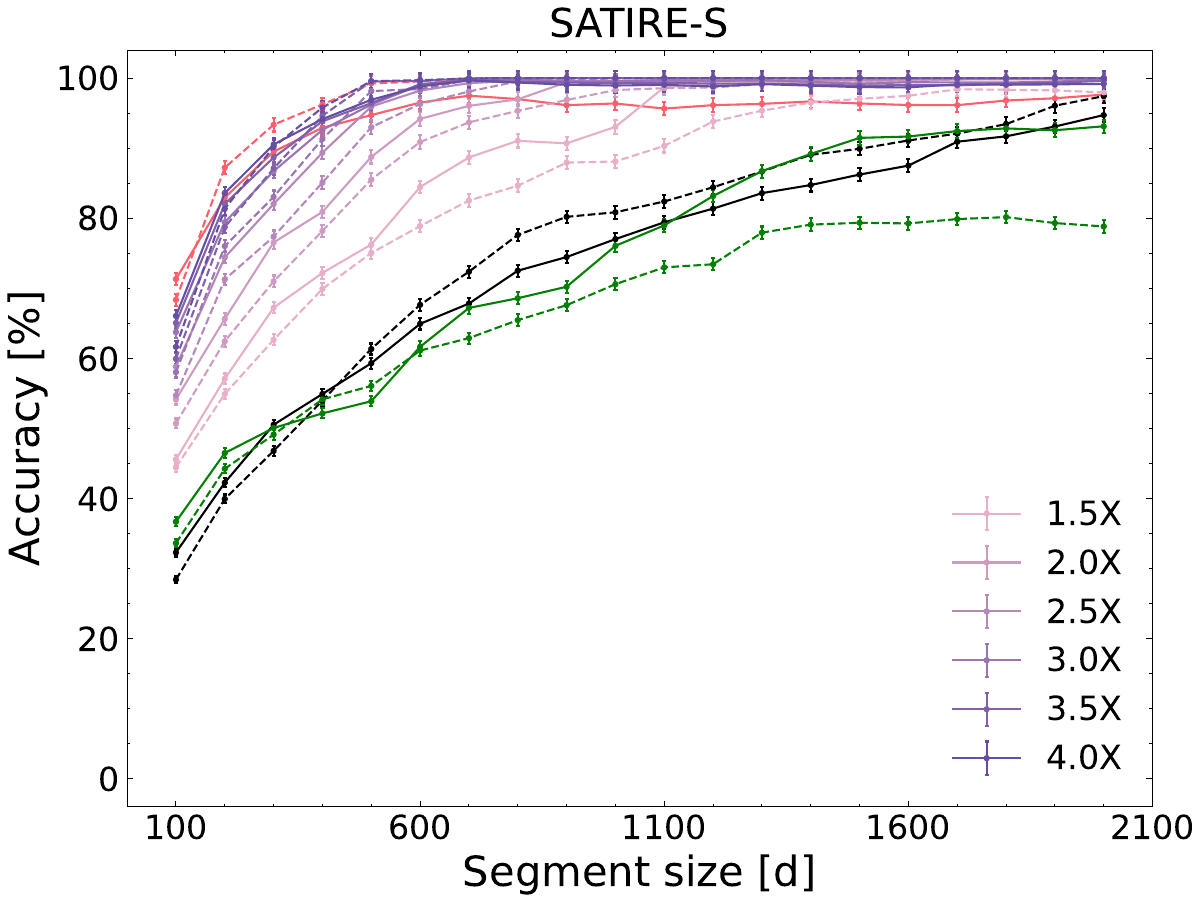}{0.533\textwidth}{}
          \fig{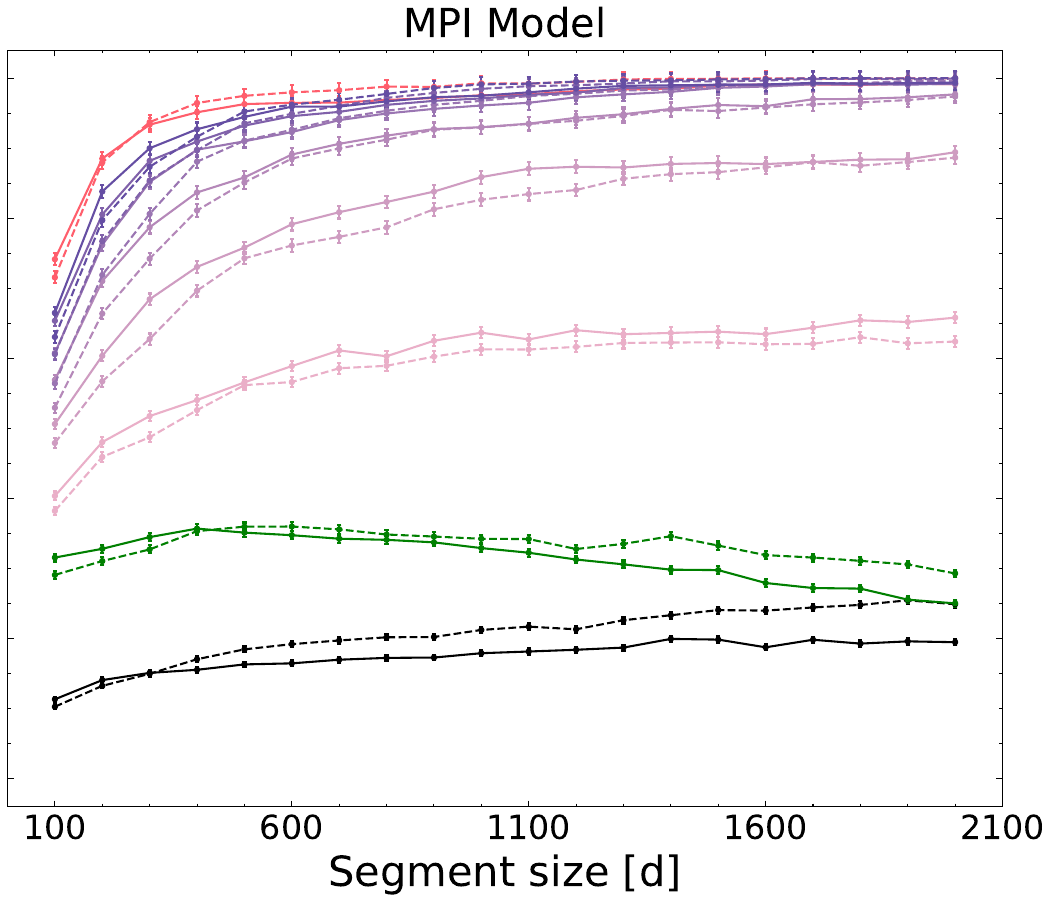}{0.467\textwidth}{}}
\caption{\textbf{Bulk ACF accuracy} as a function of Monte Carlo segment size for the SATIRE-S light curves. The results for the original facular (red), spot (black), and net (green) signals are shown. The pink-to-purple lines represent results for the net light curve where the facular component of the light curve was scaled up by between 1.5X (pink) to 4X (purple).}
\label{fig:satire_and_mpi_scaling}
\end{figure*}

\textbf{In \secr{method_comparison}, we validated that light curves generated only by sunspots are not particularly periodic while the facular light curves are extremely periodic. Despite this, the bulk accuracy on the SATIRE-S net signal is usually comparable to or even less than that of the spot signal (\figr{method_comparison} top-left panel), supporting the premise that on the timescale of rotation periods, photometric variations of Sun-like stars are dominated by spots \citep{Shapiro2016}. We next quantify by how much amplifying the magnitude of facular variations can improve the \textbf{bulk accuracy} measured from the resulting net light curve.}

Our overall \textbf{bulk accuracy} results for these different facular scalings and segment sizes in the SATIRE-S and MPI light curves are shown in \figr{satire_and_mpi_scaling} with the pink to purple colors. \textbf{For the SATIRE-S dataset}, doubling the facular signal (second lowest shaded curves) greatly improves the \textbf{bulk accuracy} compared to the original (green) net light curve, achieving essentially full accuracy for segments longer than 500 days. For very long observing segments the \textbf{bulk accuracy} for the scaled net light curves becomes as accurate as it was for the original pure facular signal. Some of the scaled net curves even performed slightly better than the pure original facular signal, though not significantly. 

\begin{figure*}[htb!]
\gridline{\fig{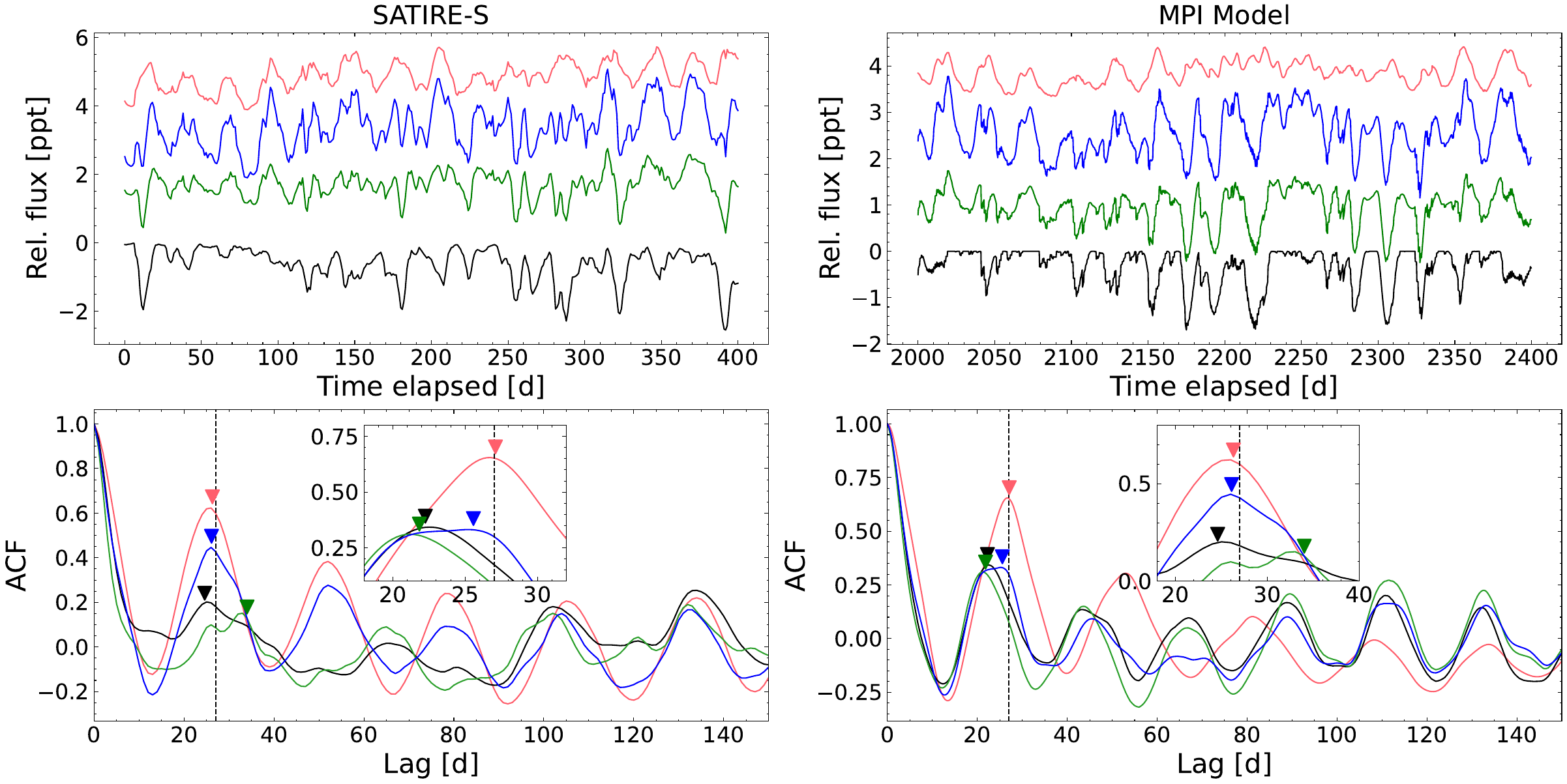}{1.0\textwidth}{}}
\caption{Examples of two ACF analyses on 400 day segments from SATIRE-S (left) and the MPI models (right). In each upper panel, the net (green), facular(red), and spot (black) light curves are shown, \textbf{vertically offset}. The net curve obtained by scaling the facular \textbf{signal} by two is shown in blue. For each ACF, the marker of corresponding color indicates the rotational period obtained by ACF-MMA. The dotted line is the solar rotational period of 27 days.}
\label{fig:mpi_and_satire_segment}
\end{figure*}

The MPI spot signal is more aperiodic than the SATIRE-S spot reconstruction in the sense that its \textbf{bulk accuracy} is substantially lower at all segment sizes. Contrary to the SATIRE-S results, the \textbf{bulk accuracy} of the \textbf{unmodified} MPI net signal decreases as the segment size increases. The accuracy of the spot signal very weakly increases. The facular signal, being very periodic \textbf{even} for the shortest segments, levels off at almost 100\% accuracy at segment sizes above 500 days. Upon scaling up the faculae, the \textbf{bulk accuracy} of the net light curves behave more and more like those from just the faculae. \textbf{For both models}, scaling up the faculae decreases the dispersion of the periods given the higher \textbf{bulk accuracy}.

The contrast between the VTSI/SATIRE-S (empirical) versus MPI (modeled) results is surprising to us. \textbf{Having already ruled out the difference in time span (2 vs. 25 solar cycles) as a possible cause}, we believe the difference in our results between the two sets of light curves is likely due to the fact that in the MPI models, the variations contributed by the spots are less periodic than the solar case actually is \textbf{for the two cycles observed that VTSI and SATIRE-S are based on}. Because the MPI curves were made to resemble solar light curves viewed under the \textit{Kepler} passband (visible light), \figr{passband} suggests that they should indeed be less accurate than results based on the VTSI. The behavior of the \textbf{bulk accuracy} with segment size, however, is not the same and suggests that something else is also different. Perhaps the flux emergence techniques used to generate the MPI models do not fully reproduce what happens on the Sun. 

In any case, we can conclude that just scaling the facular contribution up by a factor of two dramatically improves the \textbf{bulk accuracy}-inferred periods in the MPI models (\textbf{\figr{satire_and_mpi_scaling} right panel: second lowest purple curve compared to the green curve}). This effect is even stronger than in the SATIRE-S trials. This can be seen in a different modality in the right-hand panel of \figr{mpi_and_satire_segment}. Initially the ACF peak \textbf{selected by ACF-MMA} from the original net signal is very close to that from spots alone. After doubling the facular signal the resulting dominant peak is much closer to the pure facular peak, which itself is essentially at the correct solar rotational period of 27 days. Interestingly, \textbf{in the MPI example} the spot-only period is significantly more accurate than that derived from the original net signal. This is a specific example, however; \figr{satire_and_mpi_scaling} presents the overall results.

\subsection{How Long do Spots Have to Live to Induce Periodicity?}

\begin{figure*}[htb!]
\gridline{\fig{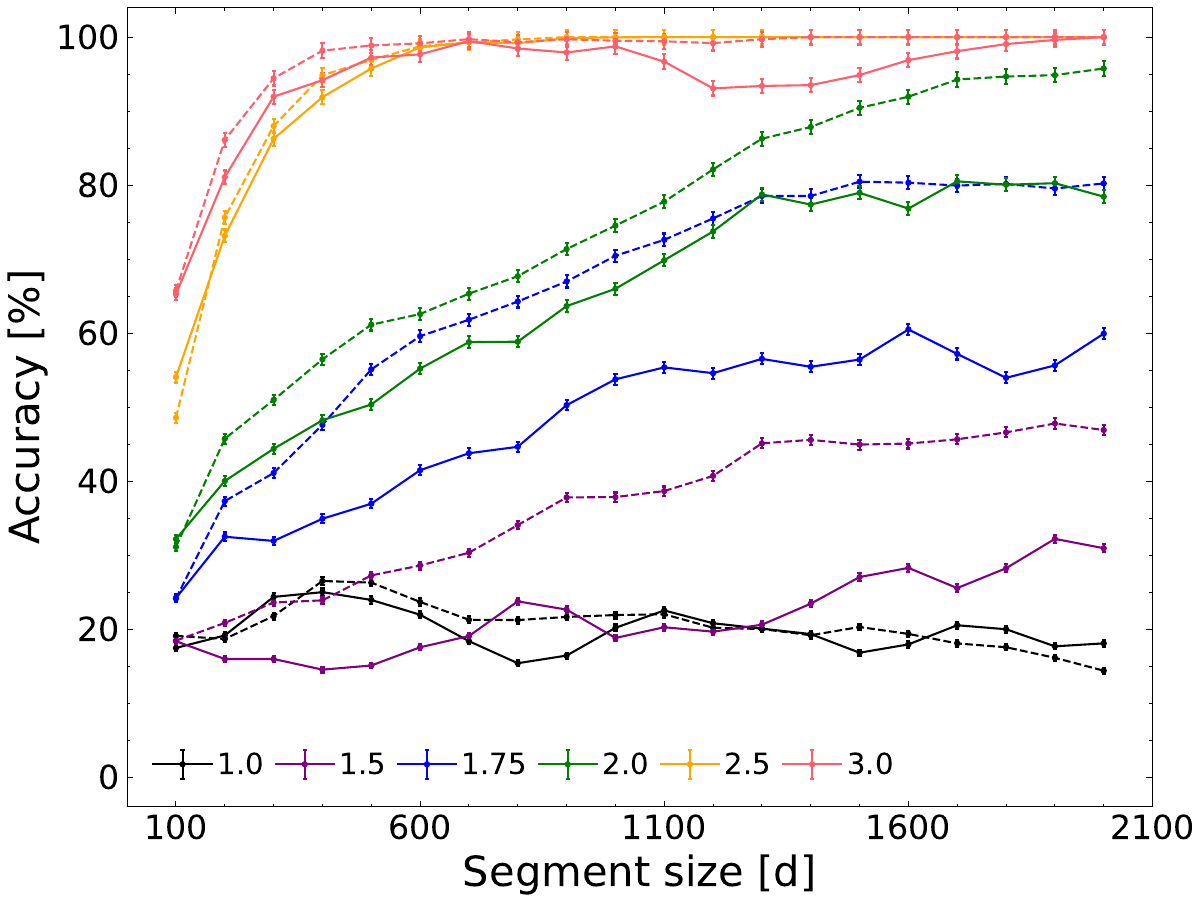}{0.6\textwidth}{}}
\caption{\textbf{Bulk ACF accuracy} as a function of Monte Carlo segment size for analytic spot models of fixed spot lifetime. Each line shows the \textbf{bulk accuracy} as a function of segment size for a specific lifetime \textbf{in rotation periods (see legend)}. \textbf{The solid and dashed lines correspond to accuracies when the segments were un\textit{Kepler}ized and \textit{Kepler}ized, respectively.}}
\label{fig:accuracy_vs_lifetime}
\end{figure*}

Given the more aperiodic nature of spot signals \textbf{that dominate the more} periodic facular signals, the short lifetime of sunspots is to blame for the inaccuracy of ACF period measurements from solar light curves. The results of our Monte Carlo trials on analytic spot models of different spot lifetimes are shown in \figr{accuracy_vs_lifetime}. The \textbf{bulk accuracy} for spots living only one rotation period is always less than 25\%, but at a lifetime of 2 solar rotation periods the \textbf{bulk accuracy} triples to over 60\% for longer segments. For \textbf{moderate lifetimes between 1.0 and 2.5 rotations}, \textit{Kepler}ization significantly improved the \textbf{bulk accuracy}, increasing the \textbf{bulk accuracy} by over 30\% at large segment sizes. This suggests that although \textit{Kepler}ization erases our knowledge of the unspotted continuum flux of the star \citep{Basri2018} and makes it more difficult to infer physical properties of starspots, it makes the light curve less subject to the effects of brighter and fainter segments interfering with each other from the perspective of the ACF. At the solar rotation period \textit{Kepler}ization is implemented over about 3 rotations (since a quarter is 90 days long).

For lifetimes of 1.5 or more rotation periods, the \textbf{bulk accuracy} increases with segment size, but the opposite occurs at shorter lifetimes. This suggests that when a signal is not particularly periodic, increasing the segment size does not improve the \textbf{bulk accuracy}. This may be because when individual segments generate ACF peaks at different periods, the composite is more likely to have spurious peaks not at harmonics of the true period. Previously we noted that for the SATIRE-S reconstruction the \textbf{bulk accuracy} for the facular, spot, and net signals each increased with segment size. In the MPI models, however, the \textbf{bulk accuracy} decreased for the spot and net signals with increasing segment size. Our results for the analytic spot model suggest that perhaps in the MPI models the spot signal is more aperiodic than for VTSI/SATIRE-S due shorter effective spot lifetimes.

\textbf{In order to test the consistency of our results, we regenerated the entire spot light curve for each lifetime multiple times, each time rerunning the Monte Carlo analysis. It turns out that the exact bulk accuracies are sensitive to the specific light curve, exhibiting a scatter on the order of $\pm20$\%. This may also be behind the less smooth behavior apparent in \figr{noise}. However, the vast majority of the bulk accuracies show the same qualitative trends against segment size, increasing with segment size for lifetimes greater than 1.5 and decreasing with segment size for lifetime 1.0. The results shown in \figr{accuracy_vs_lifetime} are representative of the retrials we conducted. Overall, although the quantitative results vary among light curves with apparently similar characteristics, the qualitative trends remain the same upon averaging.}

\subsection{\textbf{Can Faculae be Seen in} Differential Light Curves?}
\label{sec:search_for_faculae}

\begin{figure*}[htb!]
\gridline{\fig{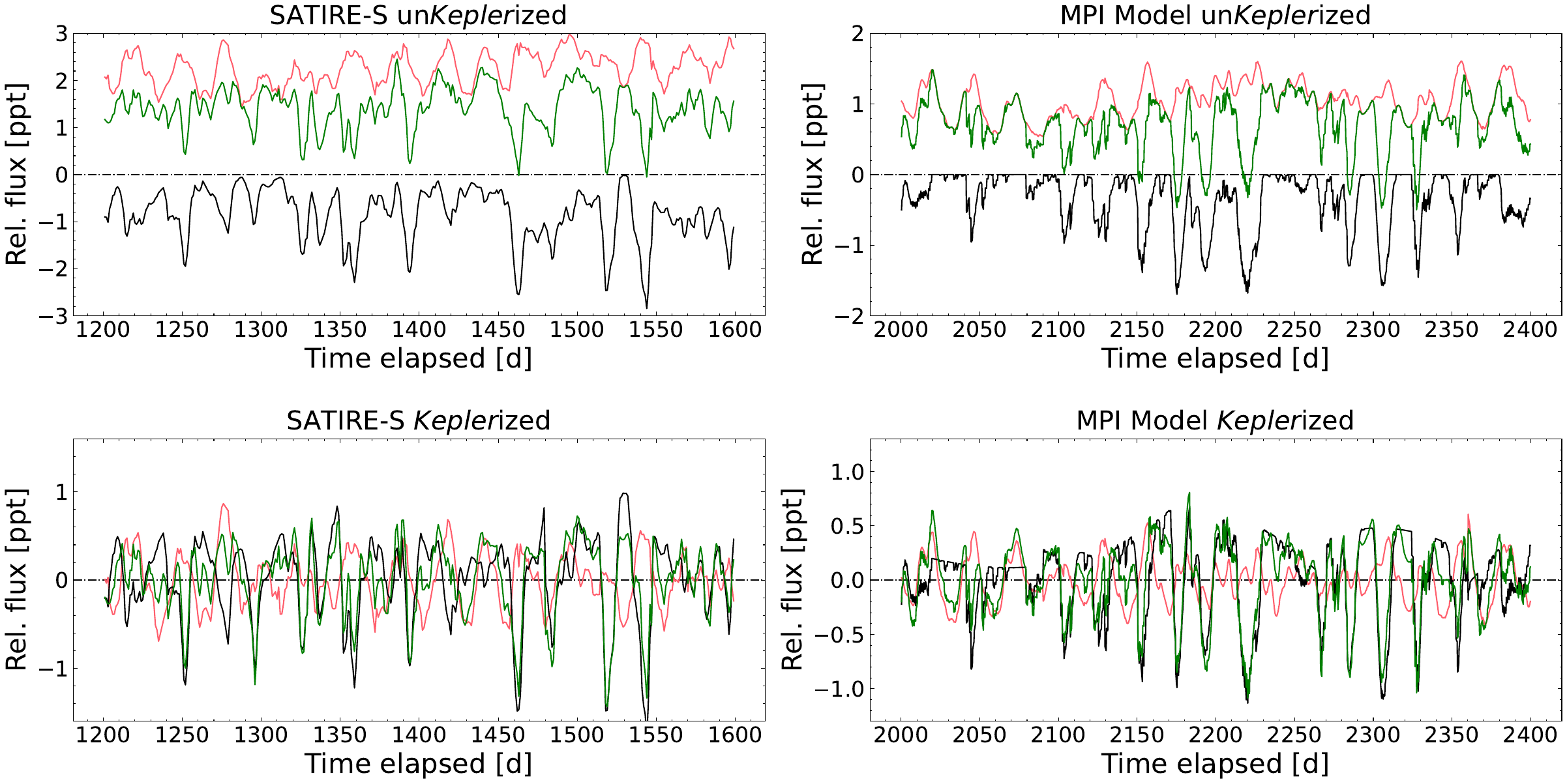}{1.0\textwidth}{}}
\caption{Illustrative examples of facular (red), spot (black), and net (green) signals at solar maximum for SATIRE-S (left panels) and the MPI model (right panels). Each set shows the un\textit{Kepler}ized (top) and \textit{Kepler}ized (bottom) versions of the same 400 day segment. }
\label{fig:search_for_faculae}
\end{figure*}

Most of what has been learned about stellar activity from the \textit{Kepler} mission is based on the photometric signal from starspots. We now ask, given a \textit{Kepler}ized stellar (net) light curve, what information can be extracted about the morphology of the facular signal? Indeed, it is not clear that anything at all has been learned about faculae from these precision differential light curves themselves. To get a better understanding of what the problem is, \figr{search_for_faculae} shows a representative example of a 400-day \textit{Kepler}ized SATIRE-S (left) and MPI (right) segment near activity maximum with its respective spot and facular components. The striking resemblance between the spot (black) and net (green) signals is immediately apparent in both the original (absolute brightness) and \textit{Kepler}ized forms. This is also consistent with the \textbf{bulk accuracy} results discussed above, which show that the net accuracies tend to track with the spot rather than facular accuracies.

Is there a possibility of deriving the facular signal from the net signal? For regions where the spot signal is weak or absent, the facular and the net strongly match before \textit{Kepler}ization. After \textit{Kepler}ization this is not necessarily true, see for example $t=2230$ to $t=2270$ in the MPI segment for a particularly extreme example. In the top panel, the red and green curve are coincident when the black curve is very close to zero, but their shapes are not the same in the bottom panel after \textit{Kepler}ization. This is because \textit{Kepler}ization does a low order fit to parts of the curve that can include regions where the spot signal is not nearly absent, so they influence regions where it is. Note that \textbf{our method of} \textit{Kepler}ization is linear; adding the individually \textit{Kepler}ized spot and facular signals produces the \textit{Kepler}ized net signal.

Not only are spotless regions very sparse when the star is active, there is no straightforward way to determine when these regions occur based on only the net signal because of lack of information about the unspotted flux \citep{Basri2018}. Even when the facular signal \textbf{varies} stronger than average, the spot signal usually is also stronger (and larger than the facular signal), effectively washing out any easily detectable facular presence from the net signal. A facular peak generally looks in the net signal just like a smaller spot with perhaps a different shape, but there is no way to know that. 

Unfortunately, we were unable to identify a method to extract any useful information about the facular signal from a net \textit{Kepler}ized light curve alone. Attempts to isolate the brighter parts of the signal did not yield fruit, nor did analysis of light curve gradients, dip locations, widths, or integrated areas. The only detectable trace of the faculae in the net light curves we found is their ability to make the net curve more periodic, and to measure how much more periodic requires separate knowledge of the spot and facular signals. It may be that the GPS method of period detection \citep{Shapiro2020} is somehow helped by the presence of faculae, but that would require a separate analysis to prove, and wouldn't by itself allow separation of the facular contribution anyway.

\section{Conclusions}

The primary purpose of this paper is to quantify the role of various observational and astrophysical effects in allowing or preventing autocorrelation function (ACF) methods from detecting stellar rotation periods in precision broadband differential photometry. These effects include varying the length, signal-to-noise, and wavelength passbands of observations, as well as their positioning within activity cycles and \textbf{the lifetimes of starspots}. We are particularly interested in the extent to which faculae, which are known to produce rather periodic photometric variability on the Sun, play a role in stellar photometry such as that gathered by the \textit{Kepler} mission. In order to study these effects on the rotational modulations of Sun-like stars we perform Monte Carlo analyses on several different datasets, measuring the \textbf{bulk accuracy} of inferred ACF rotational periods on many randomly drawn segments from solar light curves and models. Almost all of the light curves we tested are \textbf{either} based on 
\textbf{total irradiance} data and models \textbf{or the \textit{Kepler} passband}, but we also tested the effects of reducing the observational passband to various regions of the spectrum. 

In the case of the Sun, where the spot signal is irregular yet still slightly periodic, the \textbf{bulk accuracy} increases strongly with the length of the observation. Thus it is optimal to include as much data as possible for period detection. \textbf{The effect of \textit{Kepler}ization, or the removal of long term trends has mixed effects on the bulk accuracy, but generally increases it because the activity cycle trend in the ACF is removed.} We confirm that the ACF is least accurate when restricted to segments of the Sun at activity maximum, but the relation to other activity levels was weaker than expected. When observing other stars one generally does not know what phase the cycle is currently in (or if there even is a cycle), so detecting the period could be more or less difficult at another time. One might hope that stars whose variations are dominated by faculae rather than spots would be easier targets, but \textbf{even when this occurs for the Sun at low activity levels,} the variations due to faculae themselves are also rather small so that noise becomes a more important obstacle. In any case, we are unaware of any \textit{Kepler} light curves that appear to be from stars in which faculae are the primary source of measured differential variability.

We found that doubling the amplitude of the solar facular signal relative to the sunspots dramatically improves the \textbf{bulk accuracy} on the net (composite) signal. The measured periods in a Monte Carlo trial also cluster closer to the correct period (exhibit smaller dispersion) as the facular brightness variations strengthen. We quantify the effect of various levels of scaling up the facular signal. A similar helpful effect can be produced \textbf{when the inclination of the star is near 40°, where the faculae are viewed closer to the limb and appear brightest.} \textbf{And} by observing in wavelength passbands closer to the UV, the (hotter) faculae appear brighter and the spot signal is weakened. This also has the effect of allowing other signatures of magnetic active regions (heating of the chromosphere) to provide a stronger periodic signal. Indeed, the exclusion of the bluest light was part of the strategy to keep the influence of stellar activity to a minimum in {\it Kepler's} search for exoplanet transits. If one wanted to design an experiment to detect stellar rotation periods or activity levels instead, then the opposite strategy would be sensible.

In addition, the \textbf{bulk accuracy} is greatly improved when the typical lifetime of spot groups is longer than on the Sun. Spots that live for two rotations instead of one improve the accuracy of period determinations by more than twice. Using our analytic model we found that spot lifetimes of 2.5 rotations or more produce very accurate ACF measurements, with almost 100\% accuracy for segment lengths over 1000 days. It is likely that almost all the stars with periods found by ACF methods in \textit{Kepler} light curves have longer spot lifetimes than the Sun. Those that show photometric ranges larger than the Sun's are particularly likely to be spot-dominated, since the effect of spots on differential light curves grows faster than the effect of faculae as the magnetic activity increases.

We were not able to find any detectable trace of faculae in differential broadband or total irradiance light curves other than their tendency to increase the \textbf{bulk accuracy}. The primary effect of faculae in stars like the Sun is to increase their absolute brightness, so absolute photometry is required to observe that. Perhaps techniques to fit individual spots and faculae, for example Gaussian processes, could more quantitatively detect the manifestation of faculae. That is by no means certain, however, given the spatial degeneracies of the fitting problem. Methods like the GPS (based on the shape of the power spectrum) are more robust period detectors than ACF methods for \textbf{Sun-like} light curves, presumably because they rely on somewhat different diagnostics. It might be interesting to try facular scaling exercises on methods other than the ACF to see how they respond to more dominant faculae. For now, we conclude that faculae are almost irrelevant to the interpretation of any purely differential stellar photometry.

\textbf{We would like to thank the anonymous referee for providing insightful comments and constructive feedback}. Our work would not have been possible without the successful execution of the SoHO and SDO space missions. We thank members of the \textit{Max Planck Institute for Solar System Research} group for data and useful detailed discussions. GB also acknowledges the many fruitul discussions on this topic during the two ISSI meetings on ``Linking Solar and Stellar Variabilities" in Bern in 2019 and 2022. 

\facility{\textit{\textit{Kepler}}}.
\software{numpy \citep{numpy}, matplotlib \citep{matplotlib}, astropy \citep{astropy2013, astropy2018}}.


\end{document}